\documentclass[journal=jctcce,manuscript=article,layout=twocolumn]{achemso}

\usepackage[colorlinks,linkcolor=blue,citecolor=blue,urlcolor=black,bookmarks=false,hypertexnames=true]{hyperref} 

\usepackage[version=3]{mhchem}
\usepackage{amssymb}
\usepackage{color}
\usepackage{braket}
\usepackage{xspace}
\usepackage{cleveref}
\usepackage{graphicx}
\usepackage{subfig}
\usepackage{threeparttable}
\usepackage{textcomp}

\usepackage{setspace}

\usepackage{array}
\newcolumntype{L}[1]{>{\raggedright\let\newline\\\arraybackslash\hspace{0pt}}m{#1}}
\newcolumntype{C}[1]{>{\centering\let\newline\\\arraybackslash\hspace{0pt}}m{#1}}
\newcolumntype{R}[1]{>{\raggedleft\let\newline\\\arraybackslash\hspace{0pt}}m{#1}}

\makeatletter
\let\l@addto@macro\relax
\makeatother
\usepackage[fontsize=11pt]{scrextend}

\let\oldmaketitle\maketitle
\let\maketitle\relax

\newcommand{\h}[2]{h_{{#1}}^{{#2}}}
\newcommand{\f}[2]{f_{{#1}}^{{#2}}}
\renewcommand{\d}[2]{\delta_{{#1}}^{{#2}}}
\renewcommand{\v}[2]{{v}_{{#1}}^{{#2}}}

\renewcommand{\c}[1]{a^\dagger_{#1}}
\renewcommand{\a}[1]{a_{#1}}
\newcommand{\pdm}[2]{\gamma_{{#1}}^{{#2}}}

\newcommand{\e}[1]{\ensuremath{\varepsilon_{#1}}}

\newcommand*{\degree}{\ensuremath{^\circ}\xspace}

\newcommand*{\eh}{\ensuremath{E_\mathrm{h}}\xspace}
\newcommand*{\bohr}{\ensuremath{a_0}\xspace}
\newcommand{\angstrom}{\mbox{\normalfont\AA}\xspace}

\newcommand*{\mae}{$\Delta_{\mathrm{MAE}}$\xspace}
\newcommand*{\std}{$\Delta_{\mathrm{STD}}$\xspace}

\crefname{figure}{Figure}{Figures}
\crefname{table}{Table}{Tables}
\crefname{equation}{Eq.}{Eqs.}
\crefname{section}{Section}{Sections}
\crefname{subsection}{Section}{Sections}

\author{Koushik Chatterjee}
\affiliation{%
     Department of Chemistry and Biochemistry,
     The Ohio State University,
     Columbus, Ohio 43210, United States
}
 \author{Alexander Yu.\ Sokolov}
 \email{sokolov.8@osu.edu}
 \affiliation{%
     Department of Chemistry and Biochemistry,
     The Ohio State University,
     Columbus, Ohio 43210, United States
 }

\begin{tocentry}
\includegraphics{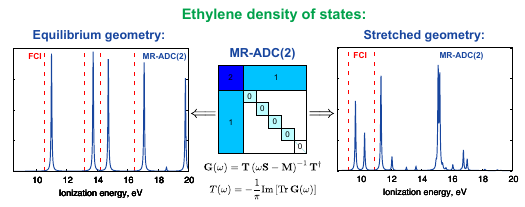}
\end{tocentry}

\newcommand*{\abstractext}{
We present a second-order formulation of multi-reference algebraic diagrammatic construction theory [Sokolov, A. Yu. {\it J. Chem. Phys.} {\bf 2018}, {\it149}, 204113] for simulating photoelectron spectra of strongly correlated systems (MR-ADC(2)). The MR-ADC(2) method uses second-order multi-reference perturbation theory (MRPT2) to efficiently obtain ionization energies and intensities for many photoelectron transitions in a single computation. In contrast to conventional MRPT2 methods, MR-ADC(2) provides information about ionization of electrons in all orbitals (i.e., core and active) and allows to compute transition intensities in straightforward and efficient way. Although equations of MR-ADC(2) depend on four-particle reduced density matrices, we demonstrate that computation of these large matrices can be completely avoided without introducing any approximations. The resulting MR-ADC(2) implementation has a lower computational scaling compared to conventional MRPT2 methods. We present results of MR-ADC(2) for photoelectron spectra of small molecules, carbon dimer, and equally-spaced hydrogen chains (\ce{H10} and \ce{H30}) and outline directions for future developments.
\vspace{0.25cm}
}

\title{Second-Order Multi-Reference Algebraic Diagrammatic Construction Theory for Photoelectron Spectra of Strongly Correlated Systems}

\begin{document}

\twocolumn[
\begin{@twocolumnfalse}
\oldmaketitle
\begin{abstract}
\abstractext
\end{abstract}
\end{@twocolumnfalse}
]

\section{Introduction}
\label{sec:intro}

Recently, there has been a significant progress in increasing tractability of strong electron correlation problem. New methods enable computations of systems with a large number of strongly correlated electrons in the ground or excited electronic 
states.\cite{Olsen:1988p2185,Malmqvist:1990p5477,White:1999p4127,Legeza2008,Booth:2009p054106,Kurashige:2009p234114,Marti:2011p6750,Chan:2011p465,Wouters:2014p272,Zhang:2016p4326,Schriber:2017p5354,Holmes:2016p3674,Sharma:2017p1595,Holmes:2017p164111} 
These approaches usually start by computing a multi-configurational wavefunction that describes strong correlation in a subset of frontier (active) molecular orbitals with 
near-degeneracies.\cite{Werner:1980p2342,Werner:1981p5794,Knowles:1985p259}
The remaining (dynamic) correlation effects outside of the active orbitals are usually captured by 
multi-reference perturbation theory (MRPT),\cite{Wolinski:1987p225,Hirao:1992p374,Werner:1996p645,Finley:1998p299,Andersson:1990p5483,Andersson:1992p1218,Angeli:2001p10252,Angeli:2001p297,Angeli:2004p4043,Li:2015p2097}
configuration interaction,\cite{Buenker:1974p33,Siegbahn:1980p1647,Werner:1988p5803,Saitow:2013p044118,Saitow:2015p5120} 
or coupled cluster (CC) methods.\cite{Mukherjee:1977p955,Lindgren:1978p33,Jeziorski:1981p1668,Mahapatra:1999p6171,Evangelista:2007p024102,Datta:2011p214116,Evangelista:2011p114102,Kohn:2012p176,Datta:2012p204107,Nooijen:2014p081102,Huntington:2015p194111,Kirtman:1981p798,Hoffmann:1988p993,Yanai:2006p194106,Yanai:2007p104107,Chen:2012p014108,Li:2016p164114,Evangelista:2018p030901} 
In particular, low-order MRPT methods have been very successful at computing accurate energies of large strongly correlated systems, due to their relatively low computation cost and ability to treat large active spaces with up to $\sim$ 30 
orbitals.\cite{Kurashige:2011p094104,Kurashige:2014p174111,Guo:2016p1583,Sharma:2017p488,Yanai:2017p4829,Freitag:2017p451,Sokolov:2017p244102,Schriber:2018p6295}

Despite significant advances, application of conventional MRPT methods to a wider range of problems, such as simulating excited-state or spectroscopic properties, is hindered by a number of limitations. For example, computation of transition intensities in MRPT is not straightforward due to complexity of the underlying response equations.\cite{MacLeod:2015p051103} Another limitation is that MRPT methods do not describe electronic transitions involving orbitals outside active space that are important for simulating broadband spectra or core-level excitations in X-ray spectroscopies. Furthermore, for computations involving many electronic states of the same symmetry, MRPT methods rely on using state-averaged reference wavefunctions, which introduce dependence of their results on the number of states and weights used in state-averaging. This motivates the development of new efficient multi-reference theories that are not bound by these limitations.

We have recently proposed a multi-reference formulation of algebraic diagrammatic construction theory (MR-ADC) for simulating spectroscopic properties of strongly correlated systems.\cite{Sokolov:2018p204113} MR-ADC is a generalization of the conventional (single-reference) ADC theory proposed by Schirmer in 1982.\cite{Schirmer:1982p2395} Rather than computing energies and wavefunctions of individual electronic states, in MR-ADC excitation energies and transition intensities are directly obtained from poles and residues of a retarded propagator approximated using multi-reference perturbation theory. In contrast to conventional MRPT, MR-ADC describes electronic transitions involving all orbitals (i.e., core, active, and external), enables simulations of various spectroscopic processes (e.g., ionization or two-photon excitation), and provides direct access to spectral properties. In this regard, MR-ADC is related to multi-reference propagator 
theories,\cite{Banerjee:1978p389,Yeager:1979p77,Dalgaard:1980p816,Yeager:1984p85,Graham:1991p2884,Yeager:1992p133,Nichols:1998p293,Khrustov:2002p507,HelmichParis:2019p174121} 
but has an advantage of a Hermitian eigenvalue problem and including dynamic correlation effects beyond single excitations. For electronic excitations, MR-ADC can also be considered as a low-cost alternative to multi-reference equation-of-motion (MR-EOM) theories, such as 
MR-EOM-CC,\cite{Datta:2012p204107,Nooijen:2014p081102,Huntington:2015p194111} and internally-contracted linear-response theories, such as ic-MRCC.\cite{Samanta:2014p134108}

In this work, we present a second-order formulation of MR-ADC (MR-ADC(2)) for photoelectron spectra of multi-reference systems. We begin by describing the derivation of MR-ADC(2) (\cref{sec:theory}) and discuss details of its implementation (\cref{sec:implementation}), demonstrating that it has a lower computational scaling with the number of active orbitals compared to conventional MRPT methods. Next, we describe computational details (\cref{sec:computational_details}) and test the performance of MR-ADC(2) for computing photoelectron energies and transition intensities of small molecules, carbon dimer, as well as equally-spaced hydrogen chains \ce{H10} and \ce{H30} (\cref{sec:results}). Finally, we present our conclusions (\cref{sec:conclusions}) and outline future developments.

\section{Theory}
\label{sec:theory}

\subsection{Multi-Reference Algebraic Diagrammatic Construction Theory (MR-ADC)}
\label{sec:theory:mr_adc_overview}

We begin with a brief overview of MR-ADC. In Ref.\@ \citenum{Sokolov:2018p204113}, we have described the derivation of MR-ADC using the formalism of effective Liouvillean theory.\cite{Mukherjee:1989p257} Here, we only summarize the main results. Our starting point is a general expression for the retarded propagator\cite{Fetter2003,Dickhoff2008} that describes response of a many-electron system to an external perturbation with frequency $\omega$:
\begin{align}
	\label{eq:g_munu}
	G_{\mu\nu}(\omega)
	& = G_{\mu\nu}^+(\omega) \pm G_{\mu\nu}^-(\omega) \notag\\
	& =  \bra{\Psi}q_\mu(\omega - H + E)^{-1}q^\dag_\nu\ket{\Psi} \notag \\
	&\pm \bra{\Psi}q^\dag_\nu(\omega + H - E)^{-1}q_\mu\ket{\Psi}
\end{align}
Here, $G_{\mu\nu}^+(\omega)$ and $G_{\mu\nu}^-(\omega)$ are the forward and backward components of the propagator, $\ket{\Psi}$ and $E$ are the eigenfunction and eigenvalue of the electronic Hamiltonian $H$, and the frequency $\omega \equiv \omega' + i\eta$ is written in terms of its real component ($\omega'$) and an infinitesimal imaginary number ($i\eta$). Depending on the form of operators $q^\dag_\nu$, the propagator $G_{\mu\nu}(\omega)$ can describe various spectroscopic processes. Choosing $q^\dag_\nu = \c{p}\a{q} - \braket{\Psi|\c{p}\a{q}|\Psi}$, where $\c{p}$ and $\a{p}$ are the usual creation and annihilation operators, corresponds to polarization propagator that provides information about electronic excitations in optical (e.g., UV/Vis) spectroscopy. Alternatively, a propagator with $q^\dag_\nu = \c{p}$ describes electron attachment and ionization processes. The number of creation and annihilation operators in $q^\dag_\nu$ (odd or even) determines the sign ($+$ or $-$) of the second term in \cref{eq:g_munu}.

Evaluation of the exact propagator is very expensive computationally. For this reason, many approximate
methods\cite{Goscinski:1980p385,Weiner:1980p1109,Prasad:1985p1287,Datta:1993p3632,Lowdin:1970p231,Nielsen:1980p6238,Sangfelt:1984p3976,Bak:2000p4173,Nooijen:1992p55,Nooijen:1993p15,Nooijen:1995p1681,Moszynski:2005p1109,Korona:2010p14977,Kowalski:2014p094102,Schirmer:1982p2395,Schirmer:1991p4647,Mertins:1996p2140,Schirmer:2004p11449,Schirmer:1983p1237,Schirmer:1998p4734,Trofimov:2005p144115,Dempwolff:2019p064108,Liu:2018p244110,Hedin:1965p796,Faleev:2004p126406,vanSchilfgaarde:2006p226402,Cederbaum:1975p290,VonNiessen:1984p57,Ortiz:2012p123,Georges:1996p13,Kotliar:2006p865,Phillips:2014p241101,Lan:2015p241102,Banerjee:1978p389,Yeager:1979p77,Dalgaard:1980p816,Yeager:1984p85,Graham:1991p2884,Yeager:1992p133,Nichols:1998p293,Khrustov:2002p507}
have been developed to compute $G_{\mu\nu}(\omega)$ for realistic systems. A common assumption in most of these approaches is that the eigenfunction $\ket{\Psi}$ can be well approximated by a single Slater determinant. Although this assumption significantly simplifies the underlying equations, such {\it single-reference} methods do not provide reliable results when strong correlation is important and the wavefunction $\ket{\Psi}$ becomes multi-configurational.

\begin{figure}[t!]
	\includegraphics[width=0.45\textwidth]{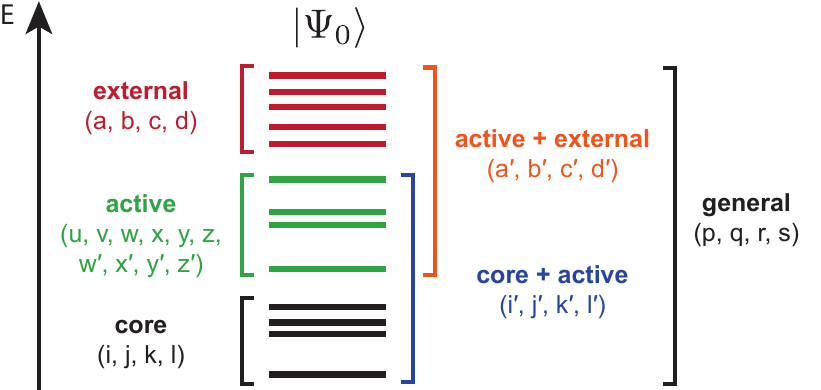}
	\captionsetup{justification=raggedright,singlelinecheck=false}
	\caption{Orbital energy diagram showing the index convention used in this work.}
	\label{fig:mo_diagram}
\end{figure}

To efficiently and accurately compute $G_{\mu\nu}(\omega)$ for strongly correlated systems, in MR-ADC we consider an expansion of \cref{eq:g_munu} using {\it multi-reference} perturbation theory, where the zeroth-order (reference) wavefunction $\ket{\Psi_0}$ is obtained by solving the complete active space configuration interaction (CASCI) or self-consistent field (CASSCF) variational problem in a set of active molecular orbitals (\cref{fig:mo_diagram}). The eigenfunction $\ket{\Psi}$ is related to $\ket{\Psi_0}$ via a unitary transformation\cite{Kirtman:1981p798,Hoffmann:1988p993,Yanai:2006p194106,Yanai:2007p104107,Chen:2012p014108,Li:2015p2097,Li:2016p164114}
\begin{align}
	\label{eq:mr_adc_wfn}
	\ket{\Psi} &= e^{A} \ket{\Psi_0} = e^{T - T^\dag} \ket{\Psi_0} , \quad T = \sum_{k=1}^N T_k  \\
	\label{eq:mr_adc_t_amplitudes}
	T_k &= \frac{1}{(k!)^2} {\sum_{i'j'a'b'\ldots}} t_{i'j'\ldots}^{a'b'\ldots} \c{a'}\c{b'}\ldots\a{j'}\a{i'}, \ t_{xy\ldots}^{wz\ldots} = 0
\end{align}
where $T$ generates all internally-contracted excitations between core, active, and external orbitals (see \cref{fig:mo_diagram} for orbital index notation). Defining the zeroth-order Hamiltonian to be the Dyall Hamiltonian\cite{Dyall:1995p4909,Angeli:2001p10252,Angeli:2001p297,Angeli:2004p4043}
\begin{align}
	\label{eq:h_dyall_general}
	H^{(0)} &\equiv C + \sum_{i} \e{i} \c{i}\a{i} + \sum_{a} \e{a} \c{a}\a{a} + H_{act} \\
	\label{eq:h_act}
	H_{act} &= \sum_{xy}(\h{x}{y} + \sum_{i} \v{xi}{yi}) \c{x} \a{y}
	+ \frac{1}{4} \sum_{xywz} \v{xy}{zw} \c{x} \c{y} \a{w} \a{z} \\
	C &= \sum_i \h{i}{i} + \frac{1}{2}\sum_{ij}\v{ij}{ij} - \sum_i  \e{i} \\
	\label{eq:f_gen}
	\f{p}{q} &= \h{p}{q} + \sum_{rs} \v{pr}{qs} \pdm{s}{r} \ , \quad \pdm{q}{p} = \braket{\Psi_0|\c{p}\a{q}|\Psi_0}
\end{align}
expressed in the basis of diagonal core and external generalized Fock operators ($\f{i}{j} = \e{i}\d{i}{j}$, $\f{a}{b} = \e{a}\d{a}{b}$), we expand the propagator in \cref{eq:g_munu} in perturbative series with respect to the perturbation $V = H - H^{(0)}$:
\begin{align}
	\label{eq:g_pt_series}
	\mathbf{G}(\omega) & = \mathbf{G}^{(0)}(\omega) + \mathbf{G}^{(1)}(\omega) + \ldots + \mathbf{G}^{(n)}(\omega) + \ldots
\end{align}
Truncating \cref{eq:g_pt_series} at the $n$th order in perturbation theory corresponds to the propagator of the MR-ADC(n) approximation.

An important property of MR-ADC (along with that of its single-reference variant)\cite{Mukherjee:1989p257} is that the forward and backward components of the propagator in \cref{eq:g_munu} are decoupled and, thus, perturbative expansion \eqref{eq:g_pt_series} can be performed for $G_{\mu\nu}^+(\omega)$ and $G_{\mu\nu}^-(\omega)$ separately. The MR-ADC(n) $G_{\mu\nu}^+(\omega)$ and $G_{\mu\nu}^-(\omega)$ contributions are expressed in the matrix form
\begin{align}
	\label{eq:Gn_matrix}
	\mathbf{G}_{\pm}(\omega) & = \mathbf{T}_{\pm} \left(\omega \mathbf{S}_{\pm} - \mathbf{M}_{\pm}\right)^{-1} \mathbf{T}_{\pm}^{\dag}
\end{align}
where $\mathbf{M}_{\pm}$, $\mathbf{T}_{\pm}$, and $\mathbf{S}_{\pm}$ are the effective Liouvillean, transition moment, and overlap matrices, respectively, each evaluated up to $n$th order in perturbation theory. The $\mathbf{M}_{\pm}$ matrix contains information about transition energies, which are obtained by solving the Hermitian generalized eigenvalue problem
\begin{align}
	\label{eq:adc_eig_problem}
	\mathbf{M}_{\pm} \mathbf{Y}_{\pm}  = \mathbf{S}_{\pm} \mathbf{Y}_{\pm} \boldsymbol{\Omega}_{\pm}
\end{align}
where $\boldsymbol{\Omega}_{\pm}$ is a diagonal matrix of eigenvalues. The eigenvectors $\mathbf{Y}_{\pm}$ are used to compute spectroscopic amplitudes
\begin{align}
	\label{eq:spec_amplitudes}
	\mathbf{X}_{\pm} = \mathbf{T}_{\pm} \mathbf{S}_{\pm}^{-1/2} \mathbf{Y}_{\pm}
\end{align}
which are related to transition intensities. Combining the eigenvalues $\boldsymbol{\Omega}_{\pm}$ and spectroscopic amplitudes $\mathbf{X}_{\pm}$, we obtain expressions for the MR-ADC(n) propagator and spectral function
\begin{align}
	\label{eq:g_mr_adc}
	\mathbf{G}_{\pm}(\omega) &= \mathbf{X}_{\pm} \left(\omega - \boldsymbol{\Omega}_{\pm}\right)^{-1}  \mathbf{X}_{\pm}^\dag \\
	\label{eq:spec_function}
	T(\omega) &= -\frac{1}{\pi} \mathrm{Im} \left[ \mathrm{Tr} \, \mathbf{G}_{\pm}(\omega) \right]
\end{align}

\subsection{Second-Order MR-ADC for Ionization Energies and Spectra}
\label{sec:theory:mr_adc_ip}

\subsubsection{Overview}
\label{sec:theory:mr_adc_ip:overview}
In this work, we consider the MR-ADC(2) approximation for photoelectron spectra, which incorporates all contributions to $\mathbf{G}(\omega)$ up to the second order in perturbation theory. A propagator of choice for the description of electron ionization processes is the backward component of the one-particle Green's function $\mathbf{G}_{-}(\omega)$, which can be defined by specifying $q^\dag_\nu = \c{p}$ in the second term of \cref{eq:g_munu}. To simplify our notation, we will drop the subscript $-$ everywhere in the equations. Thus, matrices $\mathbf{M}$, $\mathbf{T}$, and $\mathbf{S}$ will refer to the components of $\mathbf{G}_{-}(\omega)$ in \cref{eq:Gn_matrix}. Following the effective Liouvillean approach,\cite{Mukherjee:1989p257,Sokolov:2018p204113} we express the $n$th-order MR-ADC matrices as:
\begin{align}
	\label{eq:M_matrix}
	M_{\mu\nu}^{(n)} & = \sum_{klm}^{k+l+m= n} \braket{\Psi_0|[h_{\mu}^{(k)\dag}, [\tilde{H}^{(l)},h_{\nu}^{(m)}]]_{+}|\Psi_0} \\
	\label{eq:T_matrix}
	T_{\mu\nu}^{(n)} & = \sum_{kl}^{k+l=n} \braket{\Psi_0|[\tilde{q}_{\mu}^{(k)}, h_{\nu}^{(l)}]_{+}|\Psi_0} \\
	\label{eq:S_matrix}
	S_{\mu\nu}^{(n)} & = \sum_{kl}^{k+l=n} \braket{\Psi_0|[h_{\mu}^{(k)\dag}, h_{\nu}^{(l)}]_{+}|\Psi_0}
\end{align}
where $[\ldots]$ and $[\ldots]_+$ denote commutator and anticommutator, respectively. In \cref{eq:M_matrix,eq:T_matrix,eq:S_matrix}, $\tilde{H}^{(k)}$ and $\tilde{q}_{\mu}^{(k)}$ are the $k$th-order contributions to the effective Hamiltonian $\tilde{H} = e^{-A} H e^{A}$ and observable $\tilde{q}_{\mu} = e^{-A} q_{\mu} e^{A}$ operators. These contributions can be obtained by expanding $\tilde{H}$ and $\tilde{q}_{\mu}$ using the Baker--Campbell--Hausdorff (BCH) formula and collecting terms at the $k$th order. The low-order components of these operators have the form
\begin{align}
	\label{eq:H_bch_0}
	\tilde{H}^{(0)} &= H^{(0)} \\
	\label{eq:H_bch_1}
	\tilde{H}^{(1)} &= V + [H^{(0)}, A^{(1)}]  \\
	\label{eq:H_bch_2}
	\tilde{H}^{(2)} &= [H^{(0)}, A^{(2)}] + \frac{1}{2}[V + \tilde{H}^{(1)}, A^{(1)}] \\
	\label{eq:q_bch_0}
	\tilde{q}^{(0)}_\mu &= q_\mu = \a{p} \\
	\label{eq:q_bch_1}
	\tilde{q}^{(1)}_\mu &= [\a{p}, A^{(1)}] \\
	\label{eq:q_bch_2}
	\tilde{q}^{(2)}_\mu &= [\a{p}, A^{(2)}] + \frac{1}{2} [[\a{p}, A^{(1)}], A^{(1)}]
\end{align}
where $A^{(k)} = T^{(k)} - T^{(k)\dag}$ as shown in \cref{eq:mr_adc_wfn}. The operators $h_{\mu}^{(k)\dag}$ compose the $k$th-order ionization operator manifold that is used to construct a set of internally-contracted (ionized) basis states $\ket{\Psi_\mu^{(k)}} = h_{\mu}^{(k)\dag} \ket{\Psi_0}$ necessary for representing the eigenstates in \cref{eq:adc_eig_problem}.

Introducing shorthand notations\cite{Mukherjee:1989p257} for the matrix elements of arbitrary operator sets $\mathbf{A} = \{ A_\mu \}$ and $\mathbf{B} = \{ B_\mu \}$
\begin{align}
	\{\textbf{A}|\textbf{B}\} &= \braket{\Psi_0|[A_\mu,B_\nu^\dag]_+|\Psi_0} \\
	\{\textbf{A}|\tilde{\mathcal{H}}|\textbf{B}\} &= \braket{\Psi_0|[A_\mu,[\tilde{H},B_\nu^\dag]]_+|\Psi_0}
\end{align}
we express contributions to the MR-ADC(2) matrices in the following form
\begin{align}
	\label{eq:mr_adc_2_M_matrix}
	\mathbf{M} & \approx \{\mathbf{h}^{(0)\dag}|\tilde{\mathcal{H}}^{(0)}|\mathbf{h}^{(0)\dag}\} + \{\mathbf{h}^{(1)\dag}|\tilde{\mathcal{H}}^{(0)}|\mathbf{h}^{(0)\dag}\} \notag \\
	&+ \{\mathbf{h}^{(0)\dag}|\tilde{\mathcal{H}}^{(1)}|\mathbf{h}^{(0)\dag}\} + \{\mathbf{h}^{(0)\dag}|\tilde{\mathcal{H}}^{(0)}|\mathbf{h}^{(1)\dag}\} \notag \\
	&+ \{\mathbf{h}^{(1)\dag}|\tilde{\mathcal{H}}^{(1)}|\mathbf{h}^{(0)\dag}\} + \{\mathbf{h}^{(1)\dag}|\tilde{\mathcal{H}}^{(0)}|\mathbf{h}^{(1)\dag}\} \notag \\
	&+ \{\mathbf{h}^{(0)\dag}|\tilde{\mathcal{H}}^{(1)}|\mathbf{h}^{(1)\dag}\} + \{\mathbf{h}^{(0)\dag}|\tilde{\mathcal{H}}^{(2)}|\mathbf{h}^{(0)\dag}\} \\
	\label{eq:mr_adc_2_T_matrix}
	\mathbf{T} & \approx \{\mathbf{\tilde{q}}^{(0)}|\mathbf{h}^{(0)\dag}\} + \{\mathbf{\tilde{q}}^{(1)}|\mathbf{h}^{(0)\dag}\} +  \{\mathbf{\tilde{q}}^{(0)}|\mathbf{h}^{(1)\dag}\} \notag \\
	&+ \{\mathbf{\tilde{q}}^{(1)}|\mathbf{h}^{(1)\dag}\} + \{\mathbf{\tilde{q}}^{(2)}|\mathbf{h}^{(0)\dag}\}  \\
	\label{eq:mr_adc_2_S_matrix}
	\mathbf{S} & \approx \{\mathbf{h}^{(0)\dag}|\mathbf{h}^{(0)\dag}\}  + \{\mathbf{h}^{(1)\dag}|\mathbf{h}^{(0)\dag}\}  \notag \\
	&+ \{\mathbf{h}^{(0)\dag}|\mathbf{h}^{(1)\dag}\}  + \{\mathbf{h}^{(1)\dag}|\mathbf{h}^{(1)\dag}\}
\end{align}
Computing matrix elements in \cref{eq:mr_adc_2_M_matrix,eq:mr_adc_2_T_matrix,eq:mr_adc_2_S_matrix} requires solving for amplitudes of the excitation operators ($T^{(1)}$ and $T^{(2)}$) and determining the ionization operator manifolds ($h_{\mu}^{(k)\dag}$, $k = 0, 1$).

\subsubsection{Amplitudes of the Excitation Operators}
\label{sec:theory:mr_adc_ip:amplitudes}
To solve for amplitudes of the $T^{(k)}$ $(k = 1,2)$ operators, we express these operators in a general form
\begin{align}
	\label{eq:excit_op_tensor}
	T^{(k)} = \mathbf{t^{(k)}} \, \boldsymbol{\tau} =  \sum_\mu t_\mu^{(k)} \tau_\mu
\end{align}
where $t_\mu^{(k)}$ are the $k$th-order coefficients and $\tau_\mu$ are the corresponding excitation operators (\cref{eq:mr_adc_t_amplitudes}). The first-order operator $T^{(1)}$ includes up to two-body terms ($T^{(1)} = T^{(1)}_1 + T^{(1)}_2$) parametrized using three classes of single excitation and eight classes of double excitation amplitudes
\begin{align}
	\label{eq:t1_amp_tensor}
	\mathbf{t^{(1)}} = &\left\{t_{i}^{a(1)};\ t_{i}^{x(1)};\ t_{x}^{a(1)};\ t_{ij}^{ab(1)};\ t_{ij}^{ax(1)};\ t_{ix}^{ab(1)}; \right. \notag \\
	&\left. t_{ij}^{xy(1)};\ t_{xy}^{ab(1)};\ t_{ix}^{ay(1)};\ t_{ix}^{yz(1)};\ t_{xy}^{az(1)}\right\}
\end{align}
Defining $a_{q}^{p} \equiv \c{p}\a{q}$ and $a_{rs}^{pq} \equiv \c{p}\c{q}\a{s}\a{r}$, the corresponding excitation operators are
\begin{align}
	\label{eq:tau_tensor}
	\boldsymbol{\tau} = &\left\{a_{i}^{a};\ a_{i}^{x};\ a_{x}^{a};\ a_{ij}^{ab};\ a_{ij}^{ax};\ a_{ix}^{ab}; \right. \notag \\
	&\left. a_{ij}^{xy};\ a_{xy}^{ab};\ a_{ix}^{ay};\ a_{ix}^{yz};\ a_{xy}^{az}\right\}
\end{align}
To compute $\mathbf{t^{(1)}}$, we consider a system of projected linear equations
\begin{align}
	\label{eq:proj_amplitude_equations_1}
	\braket{\Psi_0|\tau^\dag_\mu\tilde{H}^{(1)}|\Psi_0} = 0
\end{align}
Using the definition of $\tilde{H}^{(1)}$ from \cref{eq:H_bch_1}, this system of equations can be expressed in the matrix form\cite{Sokolov:2018p204113}
\begin{align}
	\label{eq:proj_amplitude_equations_1_matrix}
	\mathbf{H^{(0)}} \mathbf{t^{(1)}} = - \mathbf{V^{(1)}}
\end{align}
where the zeroth-order Hamiltonian and perturbation matrix elements are defined as
\begin{align}
	\label{eq:H_zero_matrix}
	H_{\mu\nu}^{(0)} &= \braket{\Psi_0|\tau^\dag_\mu(H^{(0)} - E_0)\tau_\nu|\Psi_0} \\
	\label{eq:V_first_order_matrix}
	V_{\mu}^{(1)} &= \braket{\Psi_0|\tau^\dag_\mu V|\Psi_0}
\end{align}
and $E_0$ is the zeroth-order (reference) energy. \cref{eq:proj_amplitude_equations_1_matrix} is identical to equation that defines the first-order wavefunction in the standard Rayleigh--Schr\"odinger perturbation theory. Since $H^{(0)}$ is the Dyall Hamiltonian, the first-order MR-ADC reference wavefunction $\ket{\Psi^{(1)}} = T^{(1)} \ket{\Psi_0}$ is equivalent to the first-order wavefunction in internally-contracted second-order $N$-electron valence perturbation theory (NEVPT2).\cite{Angeli:2001p10252,Angeli:2001p297,Angeli:2004p4043} Importantly, this suggests that solutions of \cref{eq:proj_amplitude_equations_1} do not suffer from intruder-state problems, provided that $\ket{\Psi_0}$ is the ground-state reference wavefunction. The $\mathbf{t^{(1)}}$ amplitudes can be used to compute the second-order correlation correction to the reference energy
\begin{align}
	\label{eq:E_reference_second_order}
	E^{(2)} &= \braket{\Psi_0| V |\Psi^{(1)}} = \braket{\Psi_0| V T^{(1)} |\Psi_0}
\end{align}
which is equivalent to the NEVPT2 correlation energy. We note that \cref{eq:proj_amplitude_equations_1_matrix,eq:E_reference_second_order} have been recently derived in the context of perturbation expansion of internally-contracted multi-reference coupled cluster theory.\cite{Aoto:2019p2291}

Evaluating the MR-ADC(2) matrices in \cref{eq:mr_adc_2_M_matrix,eq:mr_adc_2_T_matrix} also requires semi-internal amplitudes of the second-order excitation operator $T^{(2)}$
\begin{align}
	\label{eq:t2_amp_tensor}
	\mathbf{t^{(2)}} =  \left\{t_{i}^{a(2)};\ t_{i}^{x(2)};\ t_{x}^{a(2)};\ t_{ix}^{ay(2)};\ t_{ix}^{yz(2)};\ t_{xy}^{az(2)}\right\}
\end{align}
These parameters are obtained by solving the second-order linear equations
\begin{align}
	\label{eq:proj_amplitude_equations_2_matrix}
	\mathbf{H^{(0)}} \mathbf{t^{(2)}} = - \mathbf{V^{(2)}}
\end{align}
where the matrix elements of $\mathbf{V^{(2)}}$ are defined as
\begin{align}
	\label{eq:V_second_order_matrix}
	V_{\mu}^{(2)} &= \frac{1}{2}\braket{\Psi_0|\tau^\dag_\mu [V + \tilde{H}^{(1)}, A^{(1)}]|\Psi_0}
\end{align}
\cref{eq:proj_amplitude_equations_2_matrix} is analogous to the first-order \cref{eq:proj_amplitude_equations_1_matrix} with r.h.s.\@ modified by the second-order matrix $\mathbf{V^{(2)}}$ and, thus, can be solved in a similar way. In practice, only a small number of terms in \cref{eq:mr_adc_2_M_matrix,eq:mr_adc_2_T_matrix} depend on the $\mathbf{t^{(2)}}$ amplitudes and their contributions have a very small effect on the ionization energies and spectral intensities. We will discuss solution of the first- and second-order amplitude equations in more detail in \cref{sec:implementation:amplitude_equations}.

\label{sec:theory:mr_adc_ip:ionization_operators}
\begin{figure*}[t!]
	\includegraphics[width=0.8\textwidth]{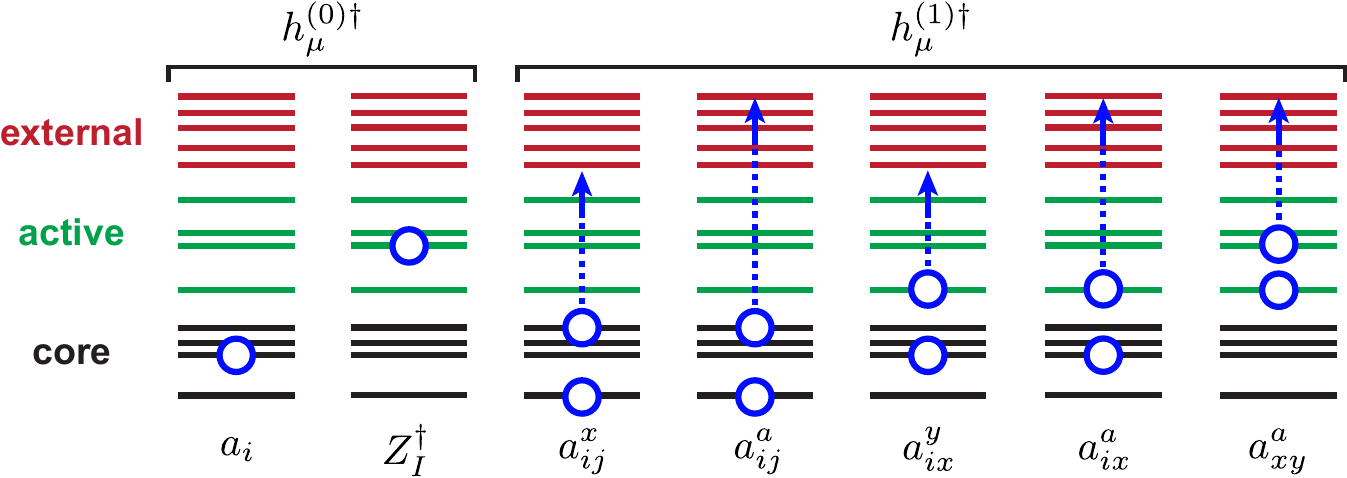}
	\captionsetup{justification=raggedright,singlelinecheck=false}
	\caption{Schematic illustration of the ionized states produced by acting the $h_{\mu}^{(0)\dag}$ and $h_{\mu}^{(1)\dag}$ operators (\cref{eq:zeroth_order_manifold,eq:first_order_manifold}) on the reference state $\ket{\Psi_0}$. The black, green, and red energy levels correspond to core, active, and external orbitals. Empty circle represents ionization and dashed line with an arrow denotes single excitation.}
	\label{fig:excitations}
\end{figure*}

\subsubsection{Ionization Operator Manifolds}
To determine the ionization operators $h_{\mu}^{(k)\dag}$ ($k = 0, 1$), we use the fact that these operators must satisfy two requirements:\cite{Mukherjee:1989p257,Sokolov:2018p204113} (i) at the $k$th order, the particle-hole rank of $h_{\mu}^{(k)\dag}$ must not exceed that of $\tilde{q}_{\mu}^{(k)\dag}$ or $\tilde{q}_{\mu}^{(k)}$ for the forward or backward components of the propagator, respectively; (ii) $h_{\mu}^{(k)\dag}$ must fulfill the vacuum annihilation condition
(VAC)\cite{Goscinski:1980p385,Weiner:1980p1109,Prasad:1985p1287,Datta:1993p3632}
with respect to the reference state, i.e.\@ $h_{\mu}^{(k)}\ket{\Psi_0} = 0$, which ensures decoupling of the forward and backward components of the propagator in \cref{eq:g_munu}.\cite{Mukherjee:1989p257,Sokolov:2018p204113}
To obtain $h_{\mu}^{(0)\dag}$, we recall that $\tilde{q}_{\mu}^{(0)} = \a{p}$, where the annihilation operator can be of three different types: $\a{i}$, $\a{x}$, or $\a{a}$ (core, active, or external). Out of these three classes, only the core operator $\a{i}$ satisfies VAC with respect to $\ket{\Psi_0}$ ($\c{i}\ket{\Psi_0} = 0$) and, thus, can be added to $h_{\mu}^{(0)\dag}$. Since $\ket{\Psi_0}$ does not contain electrons in the active space, the external operator $\a{a}$ is redundant ($\a{a}\ket{\Psi_0} = 0$) and cannot be included in $h_{\mu}^{(0)\dag}$. 
Although the active-space operator $\a{x}$ does not fulfill VAC ($\c{x}\ket{\Psi_0} \ne 0$), it can be expanded\cite{Sokolov:2018p204113} in the form $\a{x} = \sum_{I} Z^\dag_I c_{I,x}$, where $Z^\dag_I$ is a complete set of active-space eigenoperators,\cite{Freed:1977p401,Lowdin:1985p285,Kutzelnigg:1998p5578} defined as:
\begin{align}
	\label{eq:z_ketbra}
	Z^\dag_I &= \ket{\Psi_I^{N-1}}\bra{\Psi_0}
\end{align}
Here, $\ket{\Psi_I^{N-1}}$ are the CASCI states of the ionized system with $N-1$ electrons computed using the active space and one-electron basis of the reference state $\ket{\Psi_0}$. 
We note that in the context of propagator theory the configurational operators $Z^\dag_I$ were first used by Freed and Yeager\cite{Freed:1977p401} and have two important properties: they are linearly-independent and include all types of active-only ionization operators ($\a{x}$, $\c{x}\a{y}\a{z}$, $\ldots$). Incidentally, these operators also satisfy VAC with respect to $\ket{\Psi_0}$ and can be added to $h_{\mu}^{(0)\dag}$. Although we have assumed that the set of operators $Z^\dag_I$ is complete, only a subset of these operators corresponding to CASCI states in the spectral region of interest need to be included in practice. 
We summarize that the MR-ADC(2) zeroth-order manifold $h_{\mu}^{(0)\dag}$ consists of two sets of operators:
\begin{align}
\label{eq:zeroth_order_manifold}
\mathbf{h}^{(0)\dag} = \left\{\a{i}; Z^\dag_I \right\}
\end{align}

Following a similar strategy, we determine that the first-order operators $h_{\mu}^{(1)\dag}$ have a general form $a_{qr}^{p} \equiv \c{p}\a{r}\a{q}$ and can be further divided into five classes
\begin{align}
\label{eq:first_order_manifold}
\mathbf{h}^{(1)\dag} = \left\{a_{ij}^{x}; a_{ij}^{a}; a_{ix}^{y}; a_{ix}^{a}; a_{xy}^{a}\right\}
\end{align}
describing ionization in the core or active spaces accompanied by core-active, active-external, or core-external single excitations, as shown in \cref{fig:excitations}. The all-active operators $a_{zy}^{x}$ do not appear in $\mathbf{h}^{(1)\dag}$, since they are already included in the $\mathbf{h}^{(0)\dag}$ manifold by the $Z^\dag_I$ operators.

\begin{figure*}[t!]
	\subfloat[$\mathbf{M}$ matrix]{\label{fig:M_matrix}\includegraphics[width=0.30\textwidth]{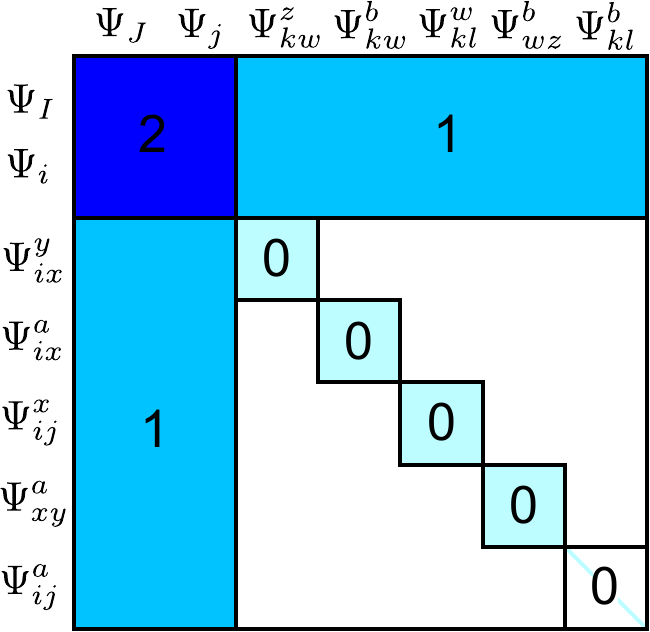}} \qquad \qquad
	\subfloat[$\mathbf{S}$ matrix]{\label{fig:S_matrix}\includegraphics[width=0.30\textwidth]{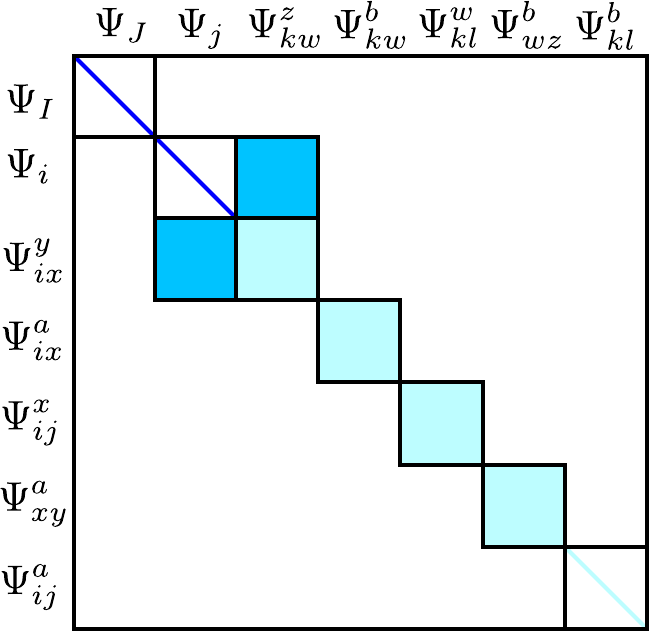}}
	\captionsetup{justification=raggedright,singlelinecheck=false}
	\caption{Structure of the effective Liouvillean ($\mathbf{M}$) and overlap ($\mathbf{S}$) matrices of MR-ADC(2) for photoelectron spectra. Non-zero matrix blocks are highlighted in color. A colored line represents a diagonal block. Numbers denote the perturbation order to which the effective Hamiltonian $\tilde{H}$ is approximated for each block. Wavefunctions $\Psi_I=Z^\dag_I\Psi_0$ and $\Psi_i=\a{i}\Psi_0$ are the CASCI and core ionized states, whereas $\Psi_{qr}^{p}=\c{p}\a{r}\a{q}\Psi_0$ are singly-excited ionized states with orbital index notation shown in \cref{fig:mo_diagram}.}
	\label{fig:M_S_matrices}
\end{figure*}

\cref{fig:M_S_matrices} illustrates perturbative structure of the MR-ADC(2) effective Liouvillean ($\mathbf{M}$) and overlap ($\mathbf{S}$) matrices. The $\{\mathbf{h}^{(0)\dag}|\tilde{\mathcal{H}}^{(k)}|\mathbf{h}^{(0)\dag}\}$ block of the $\mathbf{M}$ matrix includes all contributions up to $k$ $=$ 2, while the coupling block $\{\mathbf{h}^{(1)\dag}|\tilde{\mathcal{H}}^{(k)}|\mathbf{h}^{(0)\dag}\}$ is evaluated to first order, as given by \cref{eq:mr_adc_2_M_matrix}. In the manifold of first-order ionized states, the $\{\mathbf{h}^{(1)\dag}|\tilde{\mathcal{H}}^{(0)}|\mathbf{h}^{(1)\dag}\}$ sector is block-diagonal with non-zero elements for the $h_{\mu}^{(1)\dag}$ excitations from the same class (\cref{eq:first_order_manifold}). Overall, the general perturbative structure of the MR-ADC(2) matrices closely resembles that of non-Dyson
SR-ADC(2)\cite{Schirmer:1998p4734,Trofimov:2005p144115,Dempwolff:2019p064108}
and the two methods become equivalent in the limit of single-determinant reference wavefunction $\ket{\Psi_{0}}$.

\section{Implementation}
\label{sec:implementation}

\subsection{General Algorithm}
\label{sec:implementation:general_algorithm}
In this section, we describe a general algorithm of our MR-ADC(2) implementation for complete active space (CAS) reference wavefunctions. Although in this work we always employ the ground-state CASSCF wavefunction of a neutral system as a reference, in MR-ADC other choices of reference orbitals are possible (e.g., Hartree-Fock, state-averaged, or unrestricted natural orbitals).\cite{Bofill:1998p3637} The main steps of the MR-ADC(2) algorithm are summarized below:
\begin{enumerate}
\item Choose active space, compute the reference orbitals and CAS wavefunction $\ket{\Psi_0}$ for the neutral system with $N$ electrons.
\item Using reference orbitals, compute the CASCI energies $E_I^{N-1}$ and wavefunctions $\ket{\Psi_I^{N-1}}$ for $N_{\mathrm{CI}}$ lowest-energy states of the ionized system with $(N-1)$ electrons.
\item Compute active-space reduced density matrices (RDMs) for the reference state $\ket{\Psi_0}$, transition RDMs between $\ket{\Psi_0}$ and ionized states $\ket{\Psi_I^{N-1}}$, and transition RDMs between two ionized states $\ket{\Psi_I^{N-1}}$.
\item Solve linear amplitude \cref{eq:proj_amplitude_equations_1_matrix,eq:proj_amplitude_equations_2_matrix} to compute $\mathbf{t^{(1)}}$ and $\mathbf{t^{(2)}}$.
\item Solve the generalized eigenvalue problem \eqref{eq:adc_eig_problem} to obtain ionization energies $\boldsymbol{\Omega}$.
\item Compute spectroscopic amplitudes \eqref{eq:spec_amplitudes} and (if necessary) spectral function \eqref{eq:spec_function}.
\end{enumerate}
As discussed in \cref{sec:theory:mr_adc_ip}, the number of active-space ionized states ($N_{\mathrm{CI}}$) should be sufficiently large to include all important CASCI states in the spectral region of interest.
Implementation of the algorithm outlined above requires derivation of equations for contributions to the $\textbf{M}$, $\textbf{T}$, and $\textbf{S}$ matrices (\cref{eq:mr_adc_2_M_matrix,eq:mr_adc_2_T_matrix,eq:mr_adc_2_S_matrix}).
Although most of these contributions have compact expressions, matrix elements of the second-order effective Hamiltonian (e.g., $\{\mathbf{h}^{(0)\dag}|\tilde{\mathcal{H}}^{(2)}|\mathbf{h}^{(0)\dag}\}$) are very complicated containing $\sim$ 250-300 terms for each matrix block. Such algebraic complexity is a common feature of many internally-contracted multi-reference theories.\cite{Neuscamman:2009p124102,Datta:2012p204107,Saitow:2013p044118,MacLeod:2015p051103,Sharma:2017p488}

To speed up tedious derivation and implementation of MR-ADC(2), we have developed a Python program that automatically generates equations and code for arbitrary-order MR-ADC(n) approximation. Our code generator is a modified version of the \textsc{SecondQuantizationAlgebra} (SQA) program developed by Neuscamman and co-workers.\cite{Neuscamman:2009p124102} We use SQA to define and normal-order all active-space creation and annihilation operators in \cref{eq:mr_adc_2_M_matrix,eq:mr_adc_2_T_matrix,eq:mr_adc_2_S_matrix} with respect to the physical vacuum. Next, we additionally normal-order core creation and annihilation operators relative to the Fermi vacuum and evaluate expectation values with respect to the active-space states $\ket{\Psi_0}$ and $\ket{\Psi_I^{N-1}}$. The resulting equations, written as contractions of the one- and two-electron integrals, $\mathbf{t^{(1)}}$ and $\mathbf{t^{(2)}}$ amplitudes, and RDMs, are used to generate code and can be implemented using any available tensor contraction engine. We present working equations for all matrix elements in \cref{eq:mr_adc_2_M_matrix,eq:mr_adc_2_T_matrix,eq:mr_adc_2_S_matrix} in the Supporting Information.

In \cref{sec:implementation:amplitude_equations,sec:implementation:rdms,sec:implementation:generalized_eigenvalue_problem}, we provide more details about the solution of amplitude equations, efficient computation of terms that depend on high-order RDMs, and solution of the generalized eigenvalue problem.

\subsection{Amplitude Equations}
\label{sec:implementation:amplitude_equations}

General form of the first- and second-order amplitude equations has been discussed in \cref{sec:theory:mr_adc_ip:amplitudes}. Since the Dyall Hamiltonian (\cref{eq:h_dyall_general}) does not contain terms that couple excitations outside of the active space, its matrix representation $\mathbf{H^{(0)}}$ (\cref{eq:H_zero_matrix}) is block-diagonal and the amplitude equations \eqref{eq:proj_amplitude_equations_1_matrix} and \eqref{eq:proj_amplitude_equations_2_matrix} can be solved for each block separately. Using the standard notation for classifying excitations adopted in N-electron valence perturbation theory,\cite{Angeli:2001p10252,Angeli:2001p297,Angeli:2004p4043} operators $\boldsymbol{\tau}$ in \cref{eq:tau_tensor} are split into eight groups $\boldsymbol{\tau}^{\mathbf{[i]}}$ ($i$ $\in$ $\{0; +1; -1; +2; -2; +1'; -1'; 0'\}$), where $i$ is the number of electrons added to ($i > 0$) or removed from ($i < 0$) active space upon excitation. The operator classes with $i$ $\in$ $\{+1'; -1'; 0'\}$ are used to represent three coupled sets of single and semi-internal double excitations: $\boldsymbol{\tau}^{\mathbf{[+1']}} = \{ a_{i}^{x}; \ a_{ix}^{yz}\}$, $\boldsymbol{\tau}^{\mathbf{[-1']}} = \{ a_{x}^{a};\ a_{xy}^{az} \}$, and $\boldsymbol{\tau}^{\mathbf{[0']}} = \{ a_{i}^{a};\ a_{ix}^{ay}\}$.

Separating the $\mathbf{H^{(0)}}$, $\mathbf{t^{(1)}}$, and $\mathbf{V^{(1)}}$ matrices in \cref{eq:proj_amplitude_equations_1_matrix} into blocks according to excitation classes $\boldsymbol{\tau}^{\mathbf{[i]}}$ (denoted as $\mathbf{K^{[i]}}$, $\mathbf{t^{[i](1)}}$, and $\mathbf{V^{[i](1)}}$, respectively), we express the first-order amplitude equations in the following form
\begin{align}
	\label{eq:proj_amplitude_equations_1_subblock}
	\mathbf{K^{[i]}} \mathbf{t^{[i](1)}} = - \mathbf{V^{[i](1)}}
\end{align}
To solve \cref{eq:proj_amplitude_equations_1_subblock} for each excitation class, we consider the generalized eigenvalue problem for the matrix $\mathbf{K^{[i]}}$
\begin{align}
	\label{eq:K_eig_problem}
	\mathbf{K^{[i]}} \mathbf{Z^{[i]}} = \mathbf{S^{[i]}} \mathbf{Z^{[i]}} \boldsymbol{\epsilon}^{\mathbf{[i]}}
\end{align}
which allows to obtain expression for the first-order amplitudes\cite{Sokolov:2018p204113}
\begin{align}
	\label{eq:proj_amplitude_equations_1_subblock_solution}
	\mathbf{t^{[i](1)}} = - (\mathbf{S^{[i]}})^{-1/2}\, \mathbf{\tilde{Z}^{[i]}}\, (\boldsymbol{\epsilon}^{\mathbf{[i]}})^{-1}\, \mathbf{\tilde{Z}^{[i]\dag}} \,(\mathbf{S^{[i]}})^{-1/2} \, \mathbf{V^{[i](1)}}
\end{align}
where $K_{\mu\nu}^{[i]} = \braket{\Psi_0|\tau^{[i]\dag}_\mu(H^{(0)} - E_0)\tau^{[i]}_\nu|\Psi_0}$, $S^{[i]}_{\mu\nu} = \braket{\Psi_0|\tau^{[i]\dag}_\mu \tau^{[i]}_\nu|\Psi_0}$, and $\mathbf{\tilde{Z}^{[i]}} = (\mathbf{S^{[i]}})^{1/2}\, \mathbf{Z^{[i]}}$. Computing the $\mathbf{t^{[i](1)}}$ amplitudes in \cref{eq:proj_amplitude_equations_1_subblock_solution} requires diagonalizing $\mathbf{K^{[i]}}$ and $\mathbf{S^{[i]}}$ and removing linear dependencies corresponding to eigenvectors of $\mathbf{S^{[i]}}$ with small eigenvalues. Since the matrix elements $K_{\mu\nu}^{[i]}$ and $S^{[i]}_{\mu\nu}$ are zero when the operators $\tau^{[i]\dag}_\mu$ and $\tau^{[i]}_\nu$ do not share the same core and external indices, diagonalization of $\mathbf{K^{[i]}}$  and $\mathbf{S^{[i]}}$ can be performed very efficiently. 
For the semi-internal amplitudes $\mathbf{t^{[i](1)}}$ ($i$ $\in$ $\{+1'; -1'; 0'\}$), removing redundancies in the overlap matrix may introduce small size-consistency errors of the MR-ADC energies due to the appearance of disconnected terms in the amplitude equations that become non-zero when linear dependencies are eliminated.\cite{Sokolov:2018p204113,Hanauer:2011p204111} To restore full size-consistency of the MR-ADC energies, we use the approach developed by Hanauer and K\"ohn\cite{Hanauer:2012p131103} that removes the disconnected terms by transforming the excitation operators $\boldsymbol{\tau}^{\mathbf{[i]}}$ ($i$ $\in$ $\{+1'; -1'; 0'\}$) to a generalized normal-ordered form. We will demonstrate size-consistency of the MR-ADC(2) ionization energies in \cref{sec:results:size_consistency}.

We use \cref{eq:proj_amplitude_equations_1_subblock_solution} to compute $\mathbf{t^{[i](1)}}$ for all double ($i$ $\in$ $\{0; +1; -1; +2; -2\}$) and one class of semi-internal ($i$ $=$ $0'$) excitations. For the $\mathbf{t^{[+1'](1)}}$ and $\mathbf{t^{[-1'](1)}}$ amplitudes, diagonalization of $\mathbf{K^{[+1']}}$ and $\mathbf{K^{[-1']}}$ requires the four-particle reduced density matrix (4-RDM) of the reference state $\ket{\Psi_0}$, which is expensive to compute and store in memory for large active spaces (see \cref{sec:implementation:rdms} for details). To avoid computation of 4-RDM, we evaluate $\mathbf{t^{[+1'](1)}}$ and $\mathbf{t^{[-1'](1)}}$ using imaginary-time algorithm developed in Ref.\@ \citenum{Sokolov:2018p204113}, which employs a Laplace transform\cite{Sokolov:2016p064102,Sokolov:2017p244102} to evaluate the operator resolvent $(H^{(0)} - E_0)^{-1}$ without explicitly constructing and inverting the $\mathbf{K^{[+1']}}$ and $\mathbf{K^{[-1']}}$ matrices.

The second-order amplitude equations \eqref{eq:proj_amplitude_equations_2_matrix} need to be solved only for the semi-internal amplitudes $\mathbf{t^{[+1'](2)}}$, $\mathbf{t^{[-1'](2)}}$, and $\mathbf{t^{[0'](2)}}$ (\cref{eq:t2_amp_tensor}). Among these, only $\mathbf{t^{[+1'](2)}}$ enter equations for the $\textbf{M}$ matrix, while all three sets of semi-internal amplitudes are necessary to compute the $\textbf{T}$ matrix elements. The second-order amplitudes can be obtained in a similar way as their first-order counterparts  $\mathbf{t^{[i](1)}}$, i.e.\@ by expressing $\mathbf{t^{[i](2)}}$ in the form of \cref{eq:proj_amplitude_equations_1_subblock_solution} (with $\mathbf{V^{[i](1)}}$ replaced by $\mathbf{V^{[i](2)}}$ defined in \cref{eq:V_second_order_matrix}) or using the imaginary-time algorithm. Although solving the second-order equations is straightforward, matrix elements of the perturbation operator $\mathbf{V^{[i](2)}}$ contain $\sim$ 600 terms and are rather tedious to evaluate. On the other hand, since the primary role of $\mathbf{t^{[i](2)}}$ ($i$ $\in$ $\{+1'; -1'; 0'\}$) is to describe relaxation of the orbitals, their contributions are expected to have a small effect on the results of the MR-ADC(2) method that already incorporates orbital relaxation via the first-order amplitudes $\mathbf{t^{[i](1)}}$ and ionization operators $\mathbf{h}^{(1)\dag}$. To test this, we considered an approximation where we neglect contributions of $\mathbf{t^{[+1'](2)}}$ and $\mathbf{t^{[-1'](2)}}$ and approximate $\mathbf{t^{[0'](2)}}$ by setting $t_{ix}^{ay(2)} \approx 0$ and neglecting all terms that depend on active-space RDMs in $\mathbf{V^{[0'](2)}}$ to obtain $t_{i}^{a(2)}$ (see the Supporting Information). The resulting amplitude equations ensure that MR-ADC(2) is equivalent to SR-ADC(2) in the single-reference limit. As demonstrated in the Supporting Information, approximating the $\mathbf{t^{(2)}}$ terms has a very small effect on the MR-ADC(2) results with errors of $\le$ 0.005 eV and $\le$ $3 \times 10^{-4}$ in ionization energies and spectroscopic factors, respectively. For this reason, we adopted this approximation in our implementation of MR-ADC(2). 

\subsection{Avoiding High-Order Reduced Density Matrices}
\label{sec:implementation:rdms}
As other internally-contracted multi-reference perturbation theories, MR-ADC(2) contains terms that depend on high-order reduced density matrices (e.g., 4-RDM) in its equations. In this section, we will demonstrate that these terms can be efficiently evaluated without computing and storing 4-RDMs in memory. There are two sources of high-order RDMs in the MR-ADC(2) equations: (i) $\mathbf{t^{(1)}}$ and $\mathbf{t^{(2)}}$ amplitude equations and (ii) second-order contributions to the effective Liouvillean matrix $\mathbf{M}$. As discussed in \cref{sec:implementation:amplitude_equations}, using the imaginary-time algorithm\cite{Sokolov:2018p204113} allows to completely avoid computation of 4-RDM in the amplitude equations.

For the $\mathbf{M}$ matrix, 4-RDMs appear in expectation values of the second-order effective Hamiltonian $\tilde{\mathcal{H}}^{(2)}$ with respect to the reference ($\braket{\Psi_0|\tilde{\mathcal{H}}^{(2)}|\Psi_0}$) and ionized ($\braket{\Psi_I^{N-1}|\tilde{\mathcal{H}}^{(2)}|\Psi_J^{N-1}}$) wavefunctions. In particular, the latter matrix elements depend on transition 4-RDMs between all CASCI ionized states (e.g., $\braket{\Psi_I^{N-1}|\c{w}\c{x}\c{y}\c{z}\a{z'}\a{y'}\a{x'}\a{w'}|\Psi_J^{N-1}}$), which have a high $\mathcal{O}(N_{\mathrm{det}} N_{\mathrm{CI}}^2 N^8_{\mathrm{act}})$ computational scaling, where $N_{\mathrm{det}}$ is the dimension of CAS Hilbert space, $N_{\mathrm{CI}}$ is the number of CASCI ionized states, and $N_{\mathrm{act}}$ is the number of active orbitals. To demonstrate how to avoid computation of 4-RDMs, we consider one of the contributions to the $\braket{\Psi_I^{N-1}|\tilde{\mathcal{H}}^{(2)}|\Psi_J^{N-1}}$ matrix elements
\begin{align}
	\label{eq:avoiding_4rdm}
	\frac{1}{8} \sum_{\substack{awxyzu\\vy'w'z'}} v^{z w}_{x y} t^{x u}_{a v} t^{a y'}_{z' w'}\langle\Psi_I\lvert\c{z}\c{w}\c{u}\c{y'}\a{y}\a{v}\a{w'}\a{z'}\rvert\Psi_J\rangle
\end{align}
where we use shorthand notation for the first-order amplitudes $t_{xy}^{az(1)}\equiv t_{xy}^{az}$ and CASCI states $
\ket{\Psi_I^{N-1}} \equiv \ket{\Psi_I}$. Changing the order of creation and annihilation operators, we express \cref{eq:avoiding_4rdm} in the following form
\begin{align}
	\label{eq:avoiding_4rdm_2}
	-\frac{1}{8} \sum_{\substack{awxyzu\\vy'w'z'}} v^{z w}_{x y} t^{x u}_{a v} t^{a y'}_{z' w'}\langle\Psi_I\lvert\c{z}\c{w}\a{y}\c{u}\a{v}\c{y'}\a{w'}\a{z'}\rvert\Psi_J\rangle + \ldots
\end{align}
where the remaining terms involve contractions of transition 2- and 3-RDMs. Computing intermediate states
\begin{align}
	\label{eq:avoiding_4rdm_3}
	\rvert t^{J}_{a}\rangle &= \frac{1}{2} \sum_{y'w'z'} t^{a y'}_{z' w'} \c{y'}\a{w'}\a{z'}\rvert\Psi_J\rangle \\
	\label{eq:avoiding_4rdm_4}
	\rvert v^{I}_{x}\rangle &= \frac{1}{2} \sum_{ywz} v^{x y}_{z w} \c{y}\a{w}\a{z}\rvert\Psi_I\rangle
\end{align}
we evaluate the first term in \cref{eq:avoiding_4rdm_2} using a compact expression
\begin{align}
	\label{eq:avoiding_4rdm_5}
	-\frac{1}{2} \sum_{\substack{axuv}} t^{x u}_{a v} \langle v^{I}_{x} \lvert\c{u}\a{v}\rvert t^{J}_{a}\rangle
\end{align}
Using \cref{eq:avoiding_4rdm_3,eq:avoiding_4rdm_4,eq:avoiding_4rdm_5} allows us to significantly lower the cost of computing transition 4-RDM terms from $\mathcal{O}(N_{\mathrm{det}} N_{\mathrm{CI}}^2 N^8_{\mathrm{act}})$ to $\mathcal{O}(N_{\mathrm{det}} N_{\mathrm{CI}}^2 N^3_{\mathrm{act}} N_{\mathrm{ext}})$, where $N_{\mathrm{ext}}$ is the number of external orbitals. We use the same technique to efficiently evaluate all 4-RDM terms that appear in the $\braket{\Psi_I^{N-1}|\tilde{\mathcal{H}}^{(2)}|\Psi_J^{N-1}}$ and $\braket{\Psi_0|\tilde{\mathcal{H}}^{(2)}|\Psi_0}$ matrix elements. We note that similar techniques have been used to avoid computation of 4-RDM in implementations of complete active space second-order perturbation theory (CASPT2) and NEVPT2 in combination with matrix product state wavefunctions.\cite{Sokolov:2016p064102,Wouters:2016p054120,Sokolov:2017p244102}

The $\mathbf{M}$ matrix elements also depend on transition RDMs of the form $\braket{\Psi_0|\c{w}\c{x}\c{y}\c{z}\a{z'}\a{y'}\a{x'}|\Psi_I^{N-1}}$, which we denote as 3.5-RDMs. These RDMs contribute to the second-order matrix elements $\braket{\Psi_0|\c{i} \tilde{\mathcal{H}}^{(2)}|\Psi_I^{N-1}}$, as well as some elements of the first-order off-diagonal blocks $\{\mathbf{h}^{(1)\dag}|\tilde{\mathcal{H}}^{(1)}|\mathbf{h}^{(0)\dag}\}$ and $\{\mathbf{h}^{(0)\dag}|\tilde{\mathcal{H}}^{(1)}|\mathbf{h}^{(1)\dag}\}$ in \cref{eq:mr_adc_2_M_matrix}. For example, a 3.5-RDM contribution to $\braket{\Psi_0|\c{i} \tilde{\mathcal{H}}^{(2)}|\Psi_I^{N-1}}$ has a form
\begin{align}
	\frac{1}{8} \sum_{\substack{awxyz\\uvu'w'}} v^{x y}_{z w} t^{i z}_{a u} t^{v u'}_{a w'} \langle\Psi_{0}\lvert\c{w}\c{u}\c{v}\c{u'}\a{y}\a{x}\a{w'}\rvert\Psi_{I}\rangle
\end{align}
To evaluate this term, we reorder creation and annihilation operators, contract $v^{x y}_{z w}$ and $t^{v u'}_{a w'}$ with $\c{x}\c{y}\a{w}\rvert\Psi_{0}\rangle$ and $\c{v}\c{u'}\a{w'}\rvert\Psi_{I}\rangle$ to form intermediate states ($\rvert v^{z} \rangle$ and $\rvert t_{I}^{a} \rangle$), and contract $t^{i z}_{a u}$ with their overlap matrix element ($\langle v^{z} \lvert\c{u}\rvert t^{a}_{I} \rangle$). As in the case of 4-RDM, using intermediate states allows to completely avoid computation and storage of 3.5-RDMs for all terms of the $\mathbf{M}$ matrix, lowering computational scaling from  $\mathcal{O}(N_{\mathrm{det}} N_{\mathrm{CI}} N^7_{\mathrm{act}})$ to $\mathcal{O}(N_{\mathrm{det}} N_{\mathrm{CI}} N^2_{\mathrm{act}} N_{\mathrm{ext}})$. 

Combining efficient algorithms for the solution of amplitude equations and evaluation of high-order RDM terms, our MR-ADC(2) implementation has $\mathcal{O}(N_{\mathrm{det}} N^2_{\mathrm{CI}} N^6_{\mathrm{act}})$ computational scaling, which is significantly lower than the $\mathcal{O}(N_{\mathrm{det}} N^8_{\mathrm{act}})$ scaling of the conventional multi-reference perturbation theories (e.g., CASPT2 or NEVPT2) with the number of active orbitals. Although the scaling of our current MR-ADC(2) algorithm originates from computing transition 3-RDMs ($\braket{\Psi_I^{N-1}|\c{w}\c{x}\c{y}\a{y'}\a{x'}\a{w'}|\Psi_J^{N-1}}$) for all ionized states, we note that using intermediate states the computational cost can be further lowered to $\mathcal{O}(N_{\mathrm{det}} N_{\mathrm{CI}} N^6_{\mathrm{act}})$. We did not take advantage of it in our present implementation.

\subsection{Solution of the Generalized Eigenvalue Problem}
\label{sec:implementation:generalized_eigenvalue_problem}
Finally, we briefly discuss solution of the MR-ADC(2) generalized eigenvalue problem in \cref{eq:adc_eig_problem}. Since the $\mathbf{M}$ and $\mathbf{S}$ matrices are computed in the non-orthogonal basis of internally-contracted ionized states, we transform the eigenvalue equation to the symmetrically-orthogonalized form
\begin{align}
	\label{eq:adc_eig_problem_orthogonal}
	\mathbf{\tilde{M}} \mathbf{\tilde{Y}}  = \mathbf{\tilde{Y}} \boldsymbol{\Omega}
\end{align}
where $\mathbf{\tilde{M}} = \mathbf{S}^{-1/2} \mathbf{M} \mathbf{S}^{-1/2}$ and $\mathbf{\tilde{Y}} = \mathbf{S}^{1/2} \mathbf{Y}$.
Here, the overlap matrix $\mathbf{S}$ contains four non-diagonal blocks corresponding to ionized states $\ket{\Psi_\mu} = \{a_{ij}^{x} \ket{\Psi_0}$; $a_{ix}^{a} \ket{\Psi_0}$; $a_{xy}^{a} \ket{\Psi_0}$; $\a{i} \ket{\Psi_0}$; $a_{ix}^{y} \ket{\Psi_0}\}$ (\cref{fig:S_matrix}). Conveniently, the $\mathbf{S}^{-1/2}$ matrix can be constructed together with the $(\mathbf{S^{[i]}})^{-1/2}$ matrices used for solution of the amplitude equations (\cref{sec:implementation:amplitude_equations}). As an example, we consider non-zero elements of $\mathbf{S}$ for $a_{ij}^{x}\ket{\Psi_0}$ that have the form $S_{ijx,ijy} = \braket{\Psi_0|a_{y}^{ij}a_{ij}^{x}|\Psi_0} = \braket{\Psi_0|\a{y}\c{x}|\Psi_0}$. These elements are equal to the $\mathbf{S^{[+1]}}$ matrix elements $S_{ijay,ijax}^{[+1]} = \braket{\Psi_0|a_{ay}^{ij}a_{ij}^{ax}|\Psi_0} = \braket{\Psi_0|\a{y}\c{x}|\Psi_0}$. Thus, by diagonalizing the density matrix $\braket{\Psi_0|\a{y}\c{x}|\Psi_0}$ and removing linearly-dependent eigenvectors corresponding to small eigenvalues ($<$ $\eta_{d}$, where $\eta_{d}$ is a user-defined truncation parameter), we simultaneously obtain elements of $(\mathbf{S^{[+1]}})^{-1/2}$ and $\mathbf{S}^{-1/2}$ for the $a_{ij}^{x}\ket{\Psi_0}$ ionized wavefunctions. Similarly, we construct $(\mathbf{S^{[-1]}})^{-1/2}$ and $(\mathbf{S^{[-2]}})^{-1/2}$ together with $\mathbf{S}^{-1/2}$ for $a_{ix}^{a}\ket{\Psi_0}$ and $a_{xy}^{a}\ket{\Psi_0}$, respectively.

For the $a_{ij}^{x}\ket{\Psi_0}$, $a_{ix}^{a}\ket{\Psi_0}$, and $a_{xy}^{a}\ket{\Psi_0}$ states, numerical instabilities due to linear dependencies are completely eliminated when using small truncation parameters ($\eta_{d}$ $\sim$ $10^{-10}$). Except for very small active spaces ($N_{\mathrm{act}} < 6$), orthogonalization of these ionized states does not require discarding any eigenvectors of the overlap matrix. The zeroth-order $\a{i} \ket{\Psi_0}$ and first-order $a_{ix}^{y} \ket{\Psi_0}$ ionized states exhibit much stronger linear dependencies in their overlap matrix. To remove these linear dependencies, we project out $\a{i} \ket{\Psi_0}$ from  $a_{ix}^{y} \ket{\Psi_0}$ using the projection approach developed by Hanauer and K\"ohn\cite{Hanauer:2011p204111} and subsequently orthogonalize $a_{ix}^{y} \ket{\Psi_0}$ between each other. Importantly, this ensures that the zeroth-order states $\a{i} \ket{\Psi_0}$, which are already orthogonal, are not affected by removing redundancies in the first-order $a_{ix}^{y} \ket{\Psi_0}$ ionization manifold. To discard linearly-dependent eigenvectors of the $a_{ix}^{y} \ket{\Psi_0}$ overlap matrix, we use a larger truncation parameter ($\eta_{s}$ $\sim$ $10^{-6}$) than the one used for other ionized states ($\eta_{d}$). 

We solve the eigenvalue problem \eqref{eq:adc_eig_problem_orthogonal} using a multi-root implementation of the Davidson algorithm,\cite{Davidson:1975p87,Liu:1978p49} which avoids storing the full $\textbf{M}$ and $\textbf{S}$ matrices, significantly reducing the memory requirements. Since the second-order block $\{\mathbf{h}^{(0)\dag}|\tilde{\mathcal{H}}^{(2)}|\mathbf{h}^{(0)\dag}\}$ of $\textbf{M}$ is small (with $(N_{\mathrm{CI}} + N_{\mathrm{act}})^2$ elements) and its computation is the most time-consuming step of the MR-ADC(2) implementation, we precompute this block, store it memory, and use it for the efficient evaluation of matrix-vector products in the Davidson procedure.

\section{Computational Details}
\label{sec:computational_details}
We implemented MR-ADC(2) for photoelectron spectra in our pilot code \textsc{Prism}, which was interfaced with \textsc{Pyscf}\cite{Sun:2018pe1340} to obtain integrals and CASCI/CASSCF reference wavefunctions. Our implementation follows the general algorithm outlined in \cref{sec:implementation:general_algorithm}. All MR-ADC(2) computations used the CASSCF reference wavefunctions with molecular orbitals optimized for the ground electronic state of each (neutral) system. To remove linear dependencies in the solution of amplitude equations and generalized eigenvalue problem, we truncated eigenvectors of the overlap matrices using two parameters: $\eta_{s}$ = $10^{-6}$ and $\eta_{d}$ = $10^{-10}$ (see \cref{sec:implementation:generalized_eigenvalue_problem} for details). 
The $\eta_{s}$ parameter was used to orthogonalize the $a_{ix}^{y} \ket{\Psi_0}$ ionized states and to compute the semi-internal $\mathbf{t^{[i](1)}}$ ($i$ $\in$ $\{+1'; -1'; 0'\}$) amplitudes  (\cref{sec:implementation:amplitude_equations}), while $\eta_{d}$ was employed for other amplitudes and ionized states. 
To efficiently compute $\mathbf{t^{[+1'](1)}}$ and $\mathbf{t^{[-1'](1)}}$, our implementation used imaginary-time algorithm,\cite{Sokolov:2018p204113,Sokolov:2016p064102,Sokolov:2017p244102} where propagation in imaginary time was performed using the embedded Runge-Kutta method that automatically determines time step based on the accuracy parameter $\Delta_{it}$.\cite{Press:2007} In all computations, we used $\Delta_{it}$ = $10^{-7}$ \eh, which allows to obtain very accurate amplitudes and reference NEVPT2 correlation energy. All MR-ADC(2) results were converged with respect to the number of CASCI ionized states ($N_{\mathrm{CI}}$). For most of the systems employed in this study, using $N_{\mathrm{CI}}$ = 20 was enough to obtain well-converged results.

We benchmarked the accuracy of MR-ADC(2) for a set of small molecules (\ce{HF}, \ce{F2}, \ce{CO}, \ce{N2}, \ce{H2O}, \ce{CS}, \ce{H2CO}, and \ce{C2H4}), carbon dimer (\ce{C2}), and hydrogen chains (\ce{H10} and \ce{H30}). For small molecules, equilibrium and stretched geometries were considered. The equilibrium structures were taken from Ref.\@ \citenum{Trofimov:2005p144115}. For diatomic molecules, the stretched geometries were obtained by increasing the bond length by a factor of two. For the \ce{H2O}, \ce{H2CO}, and \ce{C2H4} stretched geometries, we doubled the \ce{O-H}, \ce{C-O}, and \ce{C-C} bond distances, respectively. The \ce{C-C} bond length in \ce{C2} was set to 1.2425 \angstrom, which is very close to its equilibrium geometry. Unless noted otherwise, all computations employed the aug-cc-pVDZ basis set.\cite{Kendall:1992p6796} For \ce{H2CO} and \ce{C2H4}, the cc-pVDZ basis set was used for the hydrogen atoms, as employed in Ref.\@ \citenum{Trofimov:2005p144115}. We denote active spaces used in CASCI/CASSCF as ($n$e, $m$o), where $n$ is the number of active electrons and $m$ is the number of active orbitals. Active spaces of small molecules included 10 orbitals with $n$ = 8, 14, 10, 10, 8, 10, 12, and 10 active electrons for \ce{HF}, \ce{F2}, \ce{CO}, \ce{N2}, \ce{H2O}, \ce{CS}, \ce{H2CO}, and \ce{C2H4}, respectively. For \ce{C2}, the (8e, 12o) active space was used. For the hydrogen chains, we employed the (10e, 10o) active space.

The MR-ADC(2) results were compared to results of single-reference non-Dyson ADC methods (SR-ADC(2) and SR-ADC(3)),\cite{Schirmer:1998p4734,Trofimov:2005p144115,Dempwolff:2019p064108} equation-of-motion coupled cluster theory for ionization energies with single and double excitations (EOM-CCSD),\cite{Sinha:1989p544,Mukhopadhyay:1991p441,Nooijen:1992p55} quasi-degenerate strongly-contracted second-order N-electron valence perturbation theory (QD-NEVPT2),\cite{Angeli:2004p4043} as well as full configuration interaction (FCI). All methods employed the same geometries and basis sets as those used for MR-ADC(2). SR-ADC(2) and SR-ADC(3) were implemented by our group as a module in the development version of \textsc{Pyscf}. The FCI results were computed using the semistochastic heat-bath configuration interaction algorithm (SHCI) implemented in the \textsc{Dice} program.\cite{Holmes:2016p3674,Sharma:2017p1595,Holmes:2017p164111} The SHCI electronic energies were extrapolated using a linear fit according to procedure described in Ref.\@ \citenum{Holmes:2017p164111}. We estimate that errors of the computed SHCI energy differences relative to FCI do not exceed 0.03 eV.
For \ce{H2CO} and \ce{C2H4}, the $1s$ atomic orbitals of carbon and oxygen were not correlated in the SHCI computations. For all other methods, all electrons were correlated in all computations. The EOM-CCSD and QD-NEVPT2 results were obtained using \textsc{Q-Chem}\cite{qchem:44} and \textsc{Orca},\cite{Neese:2017pe1327} respectively. For the ground state of each neutral system, QD-NEVPT2 used the same active spaces and CASSCF reference wavefunctions as those employed in MR-ADC(2). The QD-NEVPT2 computations of ionized states used the state-averaged CASSCF reference wavefunctions, where state-averaging included four electronic states for each abelian subgroup irreducible representation of the full symmetry point group.

Intensities of photoelectron transitions were characterized by computing spectroscopic factors
\begin{align}
	\label{eq:spec_factors}
	P_{\mu} = \sum_{p} |X_{p,\mu}|^2
\end{align}
where $X_{p,\mu}$ are elements of the spectroscopic amplitude matrix $\mathbf{X}_{\pm}$ defined in \cref{eq:spec_amplitudes}. Spectroscopic factors in \cref{eq:spec_factors} correspond to intensities of photoelectron transitions under the approximation that only single-electron detachment contributes to the spectrum. More rigorous simulation of photoelectron intensities require computation of Dyson orbitals with explicit treatment of the wavefunction of injected free electron and will be one of the subjects of our future work.\cite{Gozem:2015p4532} 

\section{Results}
\label{sec:results}

\subsection{Size-Consistency of Energies and Properties}
\label{sec:results:size_consistency}

\begin{table*}[t!]
	\captionsetup{justification=raggedright,singlelinecheck=false}
	\caption{Size-consistency errors of the MR-ADC(2) ionization energies ($\Delta \Omega$, eV) and spectroscopic factors ($\Delta P$) for the \ce{(H2O)2} and \ce{(HF)2} systems composed of two identical monomers separated by 10000 $\angstrom$ (aug-cc-pVDZ basis set). For \ce{H2O}, $r_e$ = $r$(\ce{O-H}) = 1.0 \angstrom and $\angle$(H--O--H) = 104.5\degree. For \ce{HF}, $r_e$ = 0.917 \angstrom. The \mbox{(4e, 4o)} and \mbox{(6e, 5o)} active spaces were used for the \ce{H2O} and \ce{HF} monomer CASSCF reference wavefunctions. For dimers, \mbox{(8e, 8o)} and \mbox{(12e, 10o)} active spaces were used, respectively. The number of CASCI ionized states was set to 10 and 20, for monomers and dimers, respectively.}
	\label{tab:size_consistency}
	\setstretch{1.1}
    \begin{tabular}{L{3.5cm}L{1cm}C{2.75cm}C{2.75cm}}
        \hline
        \hline
        System 					& State 		& $\Delta \Omega$ 			& $\Delta P$ 				\\
        \hline
	\ce{(H2O)2} ($r_e$)			& $1b_{1}$ 	& $-$2.4 $\times$ $10^{-5}$	& 6.0 $\times$ $10^{-8}$		\\
							& $3a_{1}$ 	& 9.4 $\times$ $10^{-6}$		& $-$2.7 $\times$ $10^{-7}$	\\
							& $1b_{2}$ 	& 2.8 $\times$ $10^{-6}$		& 2.5 $\times$ $10^{-7}$		\\							
	\ce{(H2O)2} ($2 r_e$)		& $1b_{1}$ 	& $-$1.3 $\times$ $10^{-5}$	& 3.4 $\times$ $10^{-7}$ 		\\
							& $3a_{1}$ 	& 1.5 $\times$ $10^{-5}$		& 2.3 $\times$ $10^{-6}$		\\
							& $1b_{2}$ 	& 1.3 $\times$ $10^{-5}$		& 1.8 $\times$ $10^{-6}$		\\
	\ce{(HF)2}	 ($r_e$)			& $1\pi$ 		& 1.2 $\times$ $10^{-4}$	& $-$1.0 $\times$ $10^{-6}$	\\
							& $3\sigma$	& 5.4 $\times$ $10^{-5}$	&2.4 $\times$ $10^{-6}$		\\
	\ce{(HF)2}	 ($2 r_e$)			& $1\pi$ 		& 1.0 $\times$ $10^{-4}$	&	$-$4.5 $\times$ $10^{-6}$	\\
							& $3\sigma$	& 5.9 $\times$ $10^{-5}$	&	$-$9.1 $\times$ $10^{-7}$	\\
        \hline
        \hline
    \end{tabular}
\end{table*}

We begin by testing size-consistency of the MR-ADC(2) ionization energies and spectroscopic factors. As for single-reference ADC, the MR-ADC equations are fully connected, which guarantees size-consistency of the MR-ADC energies and transition properties. In practice, however, removing redundancies in the overlap matrix during the solution of the MR-ADC amplitude equations may result in small size-consistency errors.\cite{Sokolov:2018p204113} As we discussed in \cref{sec:implementation:amplitude_equations}, in this work we employ a technique developed by Hanauer and K\"ohn that restores size-consistency of the MR-ADC results. \cref{tab:size_consistency} shows deviations from size-consistency of the MR-ADC(2) ionization energies ($\Delta \Omega$) and spectroscopic factors ($\Delta P$) for the \ce{(H2O)2} and \ce{(HF)2} systems, each composed of two noninteracting monomers with near-equilibrium ($r_e$) and stretched geometries ($2\times r_e$). The computed size-consistency errors are very small: $\Delta \Omega$ $\sim$ $10^{-5}$ eV and $\Delta P$ $\sim$ $10^{-6}$ on average, with the largest errors of \mbox{$\Delta \Omega$ = 1.2 $\times$ $10^{-4}$} eV and \mbox{$\Delta P$ = $-$4.5 $\times$ $10^{-6}$}. These remaining errors originate from a finite time step used in the imaginary-time algorithm for solving the semi-internal amplitude equations and become increasingly smaller with a tighter $\Delta_{it}$ parameter (see \cref{sec:computational_details} for details). Overall, our numerical results demonstrate size-consistency of the MR-ADC(2) results in the present implementation.

\subsection{Small Molecules}
\label{sec:results:small_molecules}

\begin{table*}[t!]
\caption{Computed vertical ionization energies ($\Omega$, eV) and spectroscopic factors ($P$) of molecules with equilibrium geometries. See \cref{sec:computational_details} for active spaces used in the reference CASSCF computations, structural parameters, and basis sets. Also shown are mean absolute errors (\mae) and standard deviations (\std) of the results, relative to FCI.}
\label{tab:ionization_energies_re}
\setstretch{1}
\scriptsize
\centering
\begin{threeparttable}
\begin{tabular}{lcccccccccccc}
\hline
\hline
System & State & \multicolumn{2}{c}{SR-ADC(2)} & \multicolumn{2}{c}{SR-ADC(3)} & \multicolumn{2}{c}{SR-ADC(3+)\tnote{a}} & \multicolumn{2}{c}{MR-ADC(2)} & EOM-CCSD & QD-NEVPT2 & FCI\tabularnewline
 &  & $\Omega$ & $P$ & $\Omega$ & $P$ & $\Omega$ & $P$ & $\Omega$ & $P$ & $\Omega$ & $\Omega$ &$\Omega$\tabularnewline
\hline
HF 			& $1\pi$ 			& 14.41 	& 0.89 	& 16.79 	& 0.93 	& 16.41 	& 0.93 	& 16.35	& 0.93 	& 15.85 	& 16.00 	& 16.07\tabularnewline
 			& $3\sigma$ 		& 18.69 	& 0.90 	& 20.65 	& 0.94 	& 20.30 	& 0.94 	& 20.38	& 0.94 	& 19.88	& 20.04 	& 20.06\tabularnewline
F$_2$ 		& $1\pi_{g}$ 		& 13.90 	& 0.87 	& 16.03 	& 0.89 	& 15.87 	& 0.90 	& 16.55	& 0.88 	& 15.40 	& 15.38 	& 15.64\tabularnewline
 			& $1\pi_{u}$ 		& 17.06 	& 0.84 	& 19.25 	& 0.75 	& 19.11 	& 0.81 	& 19.86	& 0.80 	& 18.77	& 18.58 	& 18.83\tabularnewline
 			& $3\sigma_{g}$ 	& 20.25 	& 0.89 	& 21.26 	& 0.89 	& 21.01 	& 0.88 	& 22.08	& 0.87 	& 21.16	& 20.88 	& 21.15\tabularnewline
CO 			& $5\sigma$ 		& 13.78 	& 0.91 	& 13.57 	& 0.90 	& 13.80 	& 0.89 	& 14.07	& 0.92 	& 13.99	& 13.53 	& 13.74\tabularnewline
 			& $1\pi$ 			& 16.24 	& 0.89 	& 17.16 	& 0.90 	& 16.88 	& 0.90 	& 17.38	& 0.90 	& 16.93	& 16.75 	& 16.90\tabularnewline
 			& $4\sigma$ 		& 18.28 	& 0.85 	& 20.46 	& 0.76 	& 20.10 	& 0.79 	& 20.15	& 0.85 	& 19.67	& 19.48 	& 19.56\tabularnewline
N$_2$ 		& $3\sigma_{g}$ 	& 14.79 	& 0.88 	& 15.42 	& 0.91 	& 15.60 	& 0.91 	& 15.76	& 0.91 	& 15.43	& 15.21 	& 15.30\tabularnewline
 			& $1\pi_{u}$ 		& 16.98 	& 0.91 	& 16.60 	& 0.92 	& 16.77 	& 0.92 	& 17.33	& 0.92 	& 17.11	& 16.75 	& 16.83\tabularnewline
 			& $2\sigma_{u}$ 	& 17.96 	& 0.85 	& 18.79 	& 0.82 	& 18.93 	& 0.82 	& 19.00	& 0.83 	& 18.71	& 18.44 	& 18.50\tabularnewline
H$_2$O 		& $1b_{1}$ 		& 11.23 	& 0.89 	& 12.99 	& 0.92 	& 12.78 	& 0.92 	& 12.74	& 0.93 	& 12.38	& 12.55 	& 12.53\tabularnewline
 			& $3a_{1}$ 		& 13.53 	& 0.89 	& 15.28 	& 0.92 	& 15.08 	& 0.93 	& 15.07	& 0.93 	& 14.66	& 14.85 	& 14.81\tabularnewline
 			& $1b_{2}$ 		& 17.95 	& 0.90 	& 19.34 	& 0.93 	& 19.16 	& 0.93 	& 19.28	& 0.94 	& 18.89	& 19.05 	& 18.98\tabularnewline
CS 			& $7\sigma$ 		& 10.99 	& 0.86 	& 10.99 	& 0.85 	& 11.33 	& 0.85 	& 11.59	& 0.85 	& 11.36	& 10.95 	& 11.13\tabularnewline
 			& $2\pi$ 			& 12.84 	& 0.91 	& 12.67 	& 0.90 	& 12.66 	& 0.90 	& 13.43	& 0.91 	& 12.94	& 12.74 	& 12.83\tabularnewline
 			& $6\sigma$ 		& 16.88 	& 0.85 	& 15.53 	& 0.18 	& 15.51 	& 0.19 	& 16.83	& 0.40 	& 17.02	& 15.83 	& 15.88\tabularnewline
H$_2$CO 	& $2b_{2}$ 		& 9.46 	& 0.87 	& 11.11 	& 0.91 	& 10.87 	& 0.91 	& 11.23	& 0.92 	& 10.62	& 10.28 	& 10.72\tabularnewline
 			& $1b_{1}$ 		& 13.73 	& 0.88 	& 14.54 	& 0.88 	& 14.30 	& 0.88 	& 15.14	& 0.90 	& 14.47	& 14.07 	& 14.48\tabularnewline
 			& $5a_{1}$ 		& 14.62 	& 0.86 	& 16.61 	& 0.90 	& 16.20 	& 0.90 	& 16.70	& 0.90 	& 15.95	& 15.64 	& 16.01\tabularnewline
 			& $1b_{2}$ 		& 16.67 	& 0.88 	& 17.04 	& 0.69 	& 17.32 	& 0.65 	& 17.76	& 0.88 	& 17.21	& 16.50 	& 16.86\tabularnewline
C$_2$H$_4$ 	& $1b_{1u}$ 		& 10.14 	& 0.91 	& 10.47 	& 0.91 	& 10.46 	& 0.91 	& 11.01	& 0.90 	& 10.58	& 10.41 	& 10.58\tabularnewline
 			& $1b_{1g}$ 		& 12.79 	& 0.91 	& 13.22 	& 0.91 	& 13.19 	& 0.91 	& 13.75	& 0.92 	& 13.22	& 13.05 	& 13.21\tabularnewline
 			& $3a_{g}$ 		& 13.78 	& 0.89 	& 14.34 	& 0.91 	& 14.36 	& 0.91 	& 14.74	& 0.89 	& 14.31	& 14.12 	& 14.25\tabularnewline
 			& $1b_{2u}$ 		& 16.13 	& 0.87 	& 16.50 	& 0.74 	& 16.49 	& 0.79 	& 17.10	& 0.84 	& 16.61	& 16.35 	& 16.45\tabularnewline
\mae 		&  				& 0.83 	&  		&  0.30	&  		& 0.21 	& 		& 0.56 	&  		& 0.17	& 0.17   	& \tabularnewline
\std 			&  				& 0.68 	&  		&  0.32	&  		& 0.22 	& 		& 0.23  	&  		& 0.28	& 0.14   	& \tabularnewline
\hline
\hline
\end{tabular}
	\begin{tablenotes}
		\item[a] Non-Dyson SR-ADC(3) incorporating high-order self-energy corrections from Ref.\@ \citenum{Trofimov:2005p144115}.
	\end{tablenotes}
\end{threeparttable}
\end{table*}

\begin{figure*}[t!]
    \subfloat[]{\label{fig:mae_std_re}\includegraphics[width=0.43\textwidth]{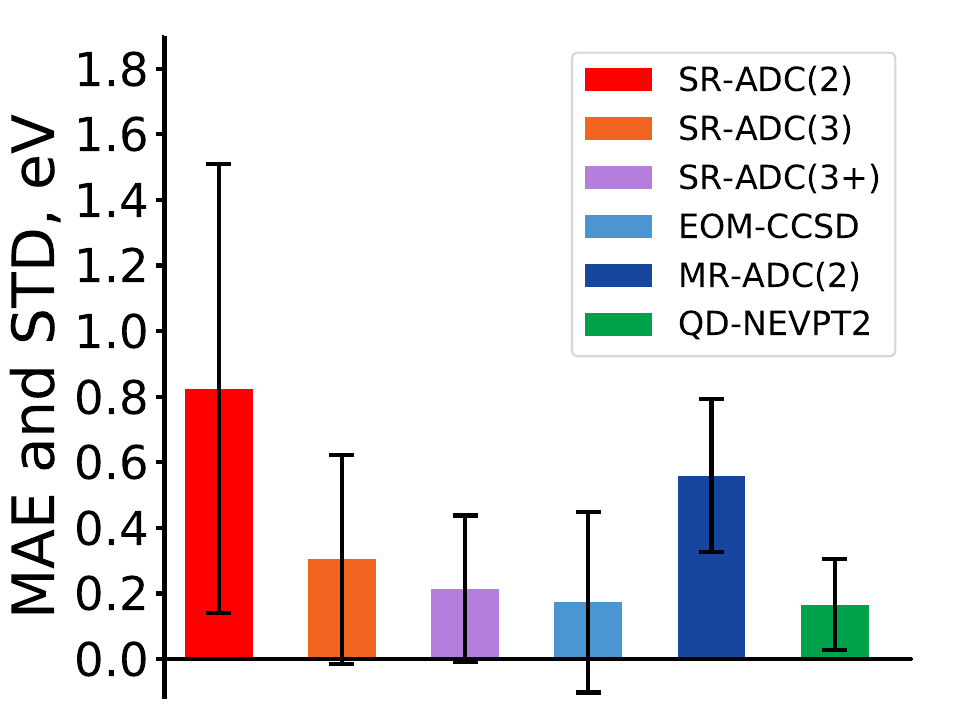}} \quad
    \subfloat[]{\label{fig:mae_std_2re}\includegraphics[width=0.43\textwidth]{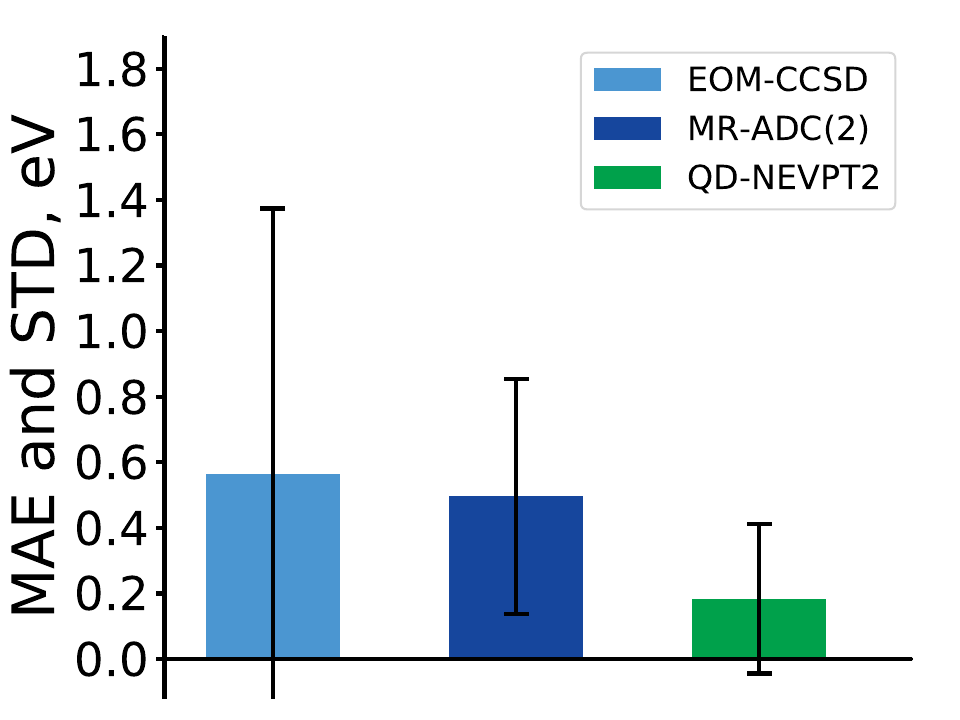}}
    \captionsetup{justification=raggedright,singlelinecheck=false}
    \caption{Mean absolute errors (MAE, eV) and standard deviations from the mean signed error (STD, eV) for vertical ionization energies of molecules with (a) equilibrium and (b) stretched geometries computed using six methods, relative to FCI (aug-cc-pVDZ basis set). The MAE value is represented as a height of each colored box, while the STD value is depicted as a radius of the black vertical bar.}
    \label{fig:mae_std}
\end{figure*}

In this section, we benchmark the MR-ADC(2) accuracy for predicting ionization energies of small molecules. \cref{tab:ionization_energies_re} compares vertical ionization energies ($\Omega$) and spectroscopic factors ($P$) of MR-ADC(2) with those obtained by single-reference non-Dyson ADC methods (SR-ADC), equation-of-motion coupled cluster theory with single and double excitations (EOM-CCSD), quasi-degenerate NEVPT2 (QD-NEVPT2), and full configuration interaction (FCI) for a set of eight molecules near their equilibrium geometries (see \cref{sec:computational_details} for computational details). In addition to strict second- and third-order SR-ADC (SR-ADC(2) and SR-ADC(3)), \cref{tab:ionization_energies_re} also presents results of SR-ADC(3) incorporating high-order self-energy corrections, reported in Ref.\@ \citenum{Trofimov:2005p144115}, which we denote as SR-ADC(3+). Out of six approximate methods, the best agreement with FCI is shown by SR-ADC(3+), EOM-CCSD, and QD-NEVPT2. These three methods produce similar mean absolute errors in vertical ionization energies (\mae $\sim$ 0.2 eV) with standard deviations from the mean signed error (\std) ranging from $\sim$ 0.15 to 0.3 eV, as illustrated in \cref{fig:mae_std_re}. The MR-ADC(2) method shows a similar \std error (0.23 eV), but a larger \mae error (0.56 eV), which is lower than \mae of SR-ADC(2) (0.83 eV), but higher than that of SR-ADC(3) (0.30 eV), indicating that including high-order effects in MR-ADC(2) improves its accuracy relative to SR-ADC(2). For all systems, the MR-ADC(2) ionization energies systematically overestimate energies computed using FCI, showing a good agreement with FCI for energy spacings between electronic states of the ionized systems (\mae of 0.11 eV and \std of 0.10 eV). The QD-NEVPT2 method shows the best agreement with FCI for energy spacings (\mae and \std of 0.03 eV), while EOM-CCSD shows larger errors compared to MR-ADC(2) (\mae = 0.16 eV, \std = 0.27 eV).
The MR-ADC(2) spectroscopic factors agree well with those computed using SR-ADC(3) and SR-ADC(3+), with two exceptions observed for the $6\sigma$ state of \ce{CS} and the $1b_{2}$ state of \ce{H2CO}. In these cases, the computed spectroscopic factors vary significantly depending on the order of the ADC approximation, suggesting that properties of these photoelectron transitions are significantly affected by electron correlation effects.

\begin{table*}[t!]
\caption{Computed vertical ionization energies ($\Omega$, eV) and spectroscopic factors ($P$) of molecules with stretched geometries. See \cref{sec:computational_details} for active spaces used in the reference CASSCF computations, structural parameters, and basis sets. Also shown are mean absolute errors (\mae) and standard deviations (\std) of the results, relative to FCI.}
\label{tab:ionization_energies_2re}
\setstretch{1}
\footnotesize
\centering
\begin{tabular}{lcccccccccc}
\hline
\hline
System & State & \multicolumn{2}{c}{SR-ADC(2)} & \multicolumn{2}{c}{SR-ADC(3)} & \multicolumn{2}{c}{MR-ADC(2)} & EOM-CCSD & QD-NEVPT2 & FCI\tabularnewline
 &  & $\Omega$ & $P$ & $\Omega$ & $P$ & $\Omega$ & $P$ & $\Omega$ & $\Omega$\tabularnewline
\hline
HF 				& $1\pi$ 			& 9.84 	& 0.77 	& 16.15 	& 0.84 	& 13.86	& 0.60	& 13.67 	& 13.61 	& 13.65\tabularnewline
				& $3\sigma$ 		& 13.30 	& 0.84 	& 14.68 	& 0.76 	& 14.98	& 0.73	& 14.76	& 14.83 	& 14.84\tabularnewline
F$_{2}$ 			& $1\pi_{g}$ 		& 10.63 	& 0.64 	& 17.55 	& 0.88 	& 18.12	& 0.74	& 16.86	& 16.81 	& 17.13\tabularnewline
				& $1\pi_{u}$ 		& 10.66 	& 0.64 	& 17.69 	& 0.89 	& 18.16	& 0.82	& 16.95	& 16.87 	& 17.19\tabularnewline
N$_{2}$ 			& $3\sigma_{g}$ 	& 15.70 	& 0.63 	& $-$2.60 	& 1.69 	& 14.00	& 0.69	& 14.36	& 13.06 	& 13.38\tabularnewline
				& $1\pi_{u}$ 		& 17.50 	& 0.55 	& $-$5.24 	& 2.16 	& 14.17	& 0.51	& 14.77	& 13.21 	& 13.49\tabularnewline
H$_{2}$O 			& $1b_{1}$ 		& 6.53 	& 0.71 	& 12.24 	& 0.66 	& 11.31	& 0.64	& 10.65	& 10.99 	& 11.07\tabularnewline
				& $3a_{1}$ 		& 10.49 	& 0.75 	& 12.78 	& 0.67 	& 13.22	& 0.67	& 12.69	& 12.99 	& 13.02\tabularnewline
				& $1b_{2}$ 		& 11.18 	& 0.75 	& 13.01 	& 0.72 	& 13.78	& 0.71	& 13.26	& 13.53 	& 13.56\tabularnewline
H$_{2}$CO 		& $2b_{2}$ 		& 10.65 	& 0.85 	& 8.31 	& 0.21 	& 11.51	& 0.39	& 9.85	& 10.24 	& 10.37\tabularnewline
 				& $1b_{1}$		& 10.69 	& 0.86 	& 8.35 	& 0.22 	& 11.21	& 0.48	& 9.66	& 10.38 	& 10.55\tabularnewline
 				& $5a_{1}$ 		& 10.60 	& 0.91 	& 10.97 	& 0.88 	& 13.16	& 0.57	& 10.97	& 12.32 	& 13.16\tabularnewline
C$_{2}$H$_{4}$ 	& $1b_{1u}$ 		& 9.37 	& 0.76 	& 6.87 	& 0.83 	& 9.69	& 0.53	& 9.41	& 9.26 	& 9.25\tabularnewline
				& $3a_{g}$ 		& 11.38 	& 0.79 	& 8.74 	& 0.91 	& 11.36	& 0.73	& 11.17	& 10.94 	& 10.93\tabularnewline
\mae				&  				& 2.70  	&  		& 3.66 	&  		& 0.50	&  		& 0.56 	& 0.18 	& \tabularnewline
\std 				&  				& 3.10  	&  		& 6.28 	&  		& 0.36	&  		& 0.81	& 0.23 	& \tabularnewline
\hline
\hline
\end{tabular}
\end{table*}

\begin{figure*}[t!]
    \subfloat[]{\label{fig:c2h4_spectrum_re}\includegraphics[width=0.45\textwidth]{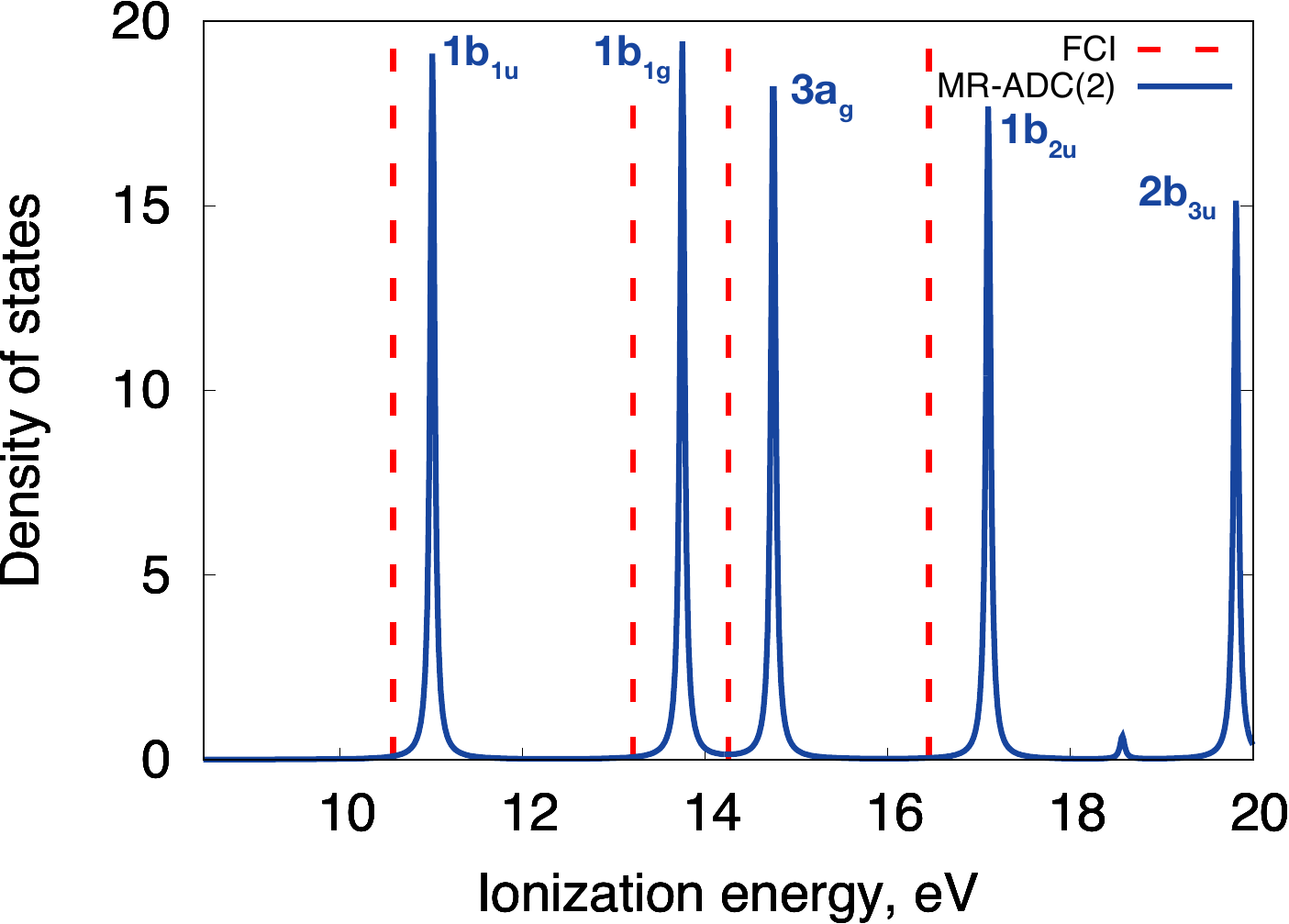}} \quad
    \subfloat[]{\label{fig:c2h4_spectrum_2re}\includegraphics[width=0.45\textwidth]{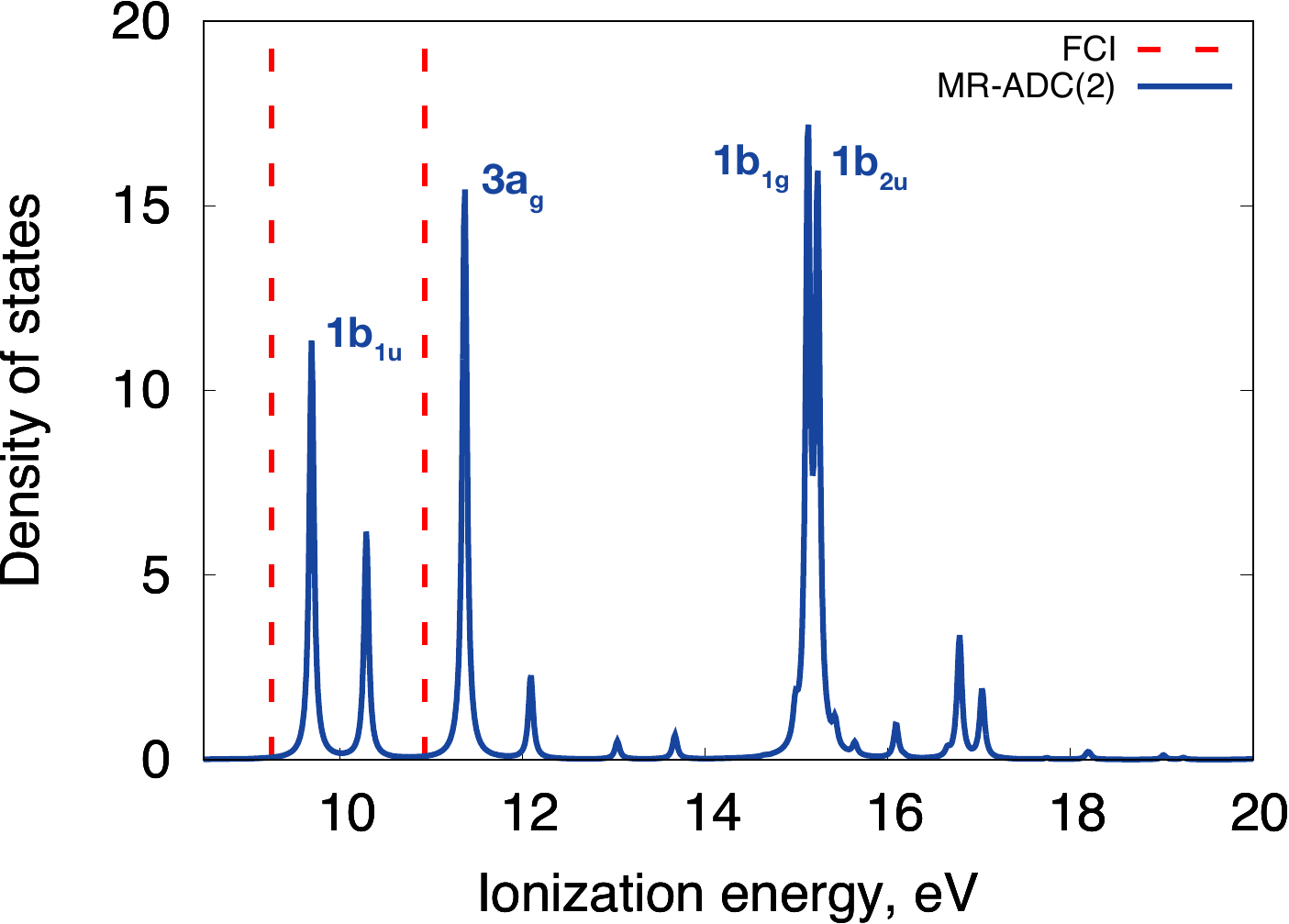}}
    \captionsetup{justification=raggedright,singlelinecheck=false}
    \caption{Simulated photoelectron spectrum of ethylene for equilibrium (a) and stretched (b) geometries computed using the MR-ADC(2) method by broadening peaks centered at vertical ionization energies with a half width of 0.03 eV (aug-cc-pVDZ basis set). Vertical dashed lines indicate FCI ionization energies for main peaks. See \cref{tab:ionization_energies_re,tab:ionization_energies_2re} for the MR-ADC(2) and FCI data.}
    \label{fig:c2h4_spectrum}
\end{figure*}

To assess performance of MR-ADC(2) when strong correlation is important, we computed its ionization energies and spectroscopic factors for molecules with stretched geometries, where at least one of the bonds is elongated by a factor of two (see \cref{sec:computational_details} for details). The MR-ADC(2) results are shown in \cref{tab:ionization_energies_2re}, along with those computed using SR-ADC(2), SR-ADC(3), EOM-CCSD, QD-NEVPT2, and FCI. Due to the difficulty of obtaining the FCI energies, we show results only for a few lowest-energy transitions of six molecules. Importance of strong electron correlation for these non-equilibrium geometries is demonstrated by the poor performance of SR-ADC(2) and SR-ADC(3), which show very large deviations from the FCI reference values with \mae $>$ 2.5 eV and \std $>$ 3 eV.  Although SR-ADC(3) shows moderate $\sim$ 0.5 eV errors for single-bond stretching in HF and \ce{F2}, these errors drastically increase when multiple bonds are elongated, leading to unphysical values of ionization energies that significantly underestimate the FCI results. EOM-CCSD significantly improves prediction of ionization energies over SR-ADC(2) and SR-ADC(3), but still exhibits large errors (\mae = 0.56 and \std = 0.81 eV, \cref{fig:mae_std_2re}). MR-ADC(2) shows performance similar to that for equilibrium geometries (\cref{tab:ionization_energies_re}), with \mae (0.50 eV) and \std (0.36 eV) smaller than the corresponding errors for the single-reference methods. The best agreement with FCI is again shown by QD-NEVPT2 with \mae = 0.18 eV and \std = 0.23 eV. As for the equilibrium geometries, the MR-ADC(2) ionization energies for stretched geometries are systematically overestimated relative to FCI, reproducing energy spacings between ionized states within 0.1 eV for all systems except  \ce{H2CO}, where $\sim$ 0.5 eV errors are observed. QD-NEVPT2 shows a similar performance to MR-ADC(2) for energy spacings with a large error of $\sim$ 0.7 eV for the difference of the $1b_{1}$ and $5a_{1}$ ionization energies of \ce{H2CO}.

\begin{table*}[t!]
\caption{Vertical ionization energies ($\Omega$, eV) and spectroscopic factors ($P$) of carbon dimer with $r$\ce{(C-C)} = 1.2425 \angstrom computed using the aug-cc-pVDZ basis set. For MR-ADC(2) and QD-NEVPT2, the CASSCF reference wavefunction was computed using the (8e, 12o) active space.}
\label{tab:c2}
\setstretch{1}
\small
\centering
\begin{threeparttable}
\begin{tabular}{lccccccc}
\hline
\hline
Configuration & State & \multicolumn{2}{c}{SR-ADC(3)} & \multicolumn{2}{c}{MR-ADC(2)} & QD-NEVPT2 & FCI\tabularnewline
  & & $\Omega$ & $P$ & $\Omega$ & $P$ & $\Omega$ & $\Omega$\tabularnewline
\hline
$(2\sigma_u)^2(1\pi_u)^3(3\sigma_g)^0$		& $1^2\Pi_u$		& 11.69 		& 0.9215 		& 12.50 	& 0.8986	& 12.28 	& 12.34\tabularnewline
$(2\sigma_u)^2(1\pi_u)^2(3\sigma_g)^1$		& $1^2\Delta_g$	& 11.17 		& 0.0002 		& 14.31 	& 0.0002	& 13.92 	& 13.94\tabularnewline
$(2\sigma_u)^2(1\pi_u)^2(3\sigma_g)^1$		& $1^2\Sigma_g^-$	& \tnote{a} 	& \tnote{a}		& 14.55 	& 0.0000	& 14.12 	& 14.15\tabularnewline
$(2\sigma_u)^2(1\pi_u)^2(3\sigma_g)^1$		& $1^2\Sigma_g^+$	& 11.43 		& 0.0004 		& 14.60 	& 0.0047	& 14.26 	& 14.29\tabularnewline
$(2\sigma_u)^1(1\pi_u)^4(3\sigma_g)^0$		& $1^2\Sigma_u^+$	& 13.95 		& 0.8738 		& 15.33	& 0.7190	& 15.07 	& 15.09\tabularnewline
$(2\sigma_u)^2(1\pi_u)^1(3\sigma_g)^2$		& $2^2\Pi_u$		& \tnote{a} 	& \tnote{a} 	& 14.77 	& 0.0183	& 15.35 	& 15.43\tabularnewline
\hline
\hline
\end{tabular}
	\begin{tablenotes}
		\item[a] State is absent in SR-ADC(3).
	\end{tablenotes}
\end{threeparttable}
\end{table*}

An important advantage of MR-ADC(2) over conventional multi-reference perturbation theories (such as QD-NEVPT2) is that it provides efficient access to spectroscopic properties. We demonstrate this by computing the photoelectron spectrum of \ce{C2H4} at equilibrium and stretched geometries in the range between 8.5 and 20 eV, shown in \cref{fig:c2h4_spectrum}. The spectrum at equilibrium geometry exhibits five very intense well-separated peaks corresponding to vertical ionizations in five highest occupied molecular orbitals. All of the computed peaks are systematically shifted by $\sim$ 0.5 eV relative to FCI.
The computed spacings between the main peaks are in a good agreement with FCI (\cref{tab:ionization_energies_re}), as well as experimental photoelectron spectrum.\cite{Branton:1970p802} At the stretched geometry, the MR-ADC(2) spectrum shows four main peaks with significantly decreased intensities, along with several satellite peaks originating from shake-up transitions that involve ionization and simultaneous excitation in the valence orbitals.

\subsection{Carbon Dimer}
\label{sec:results:carbon_dimer}

\begin{figure}[t!]
    \includegraphics[width=0.45\textwidth]{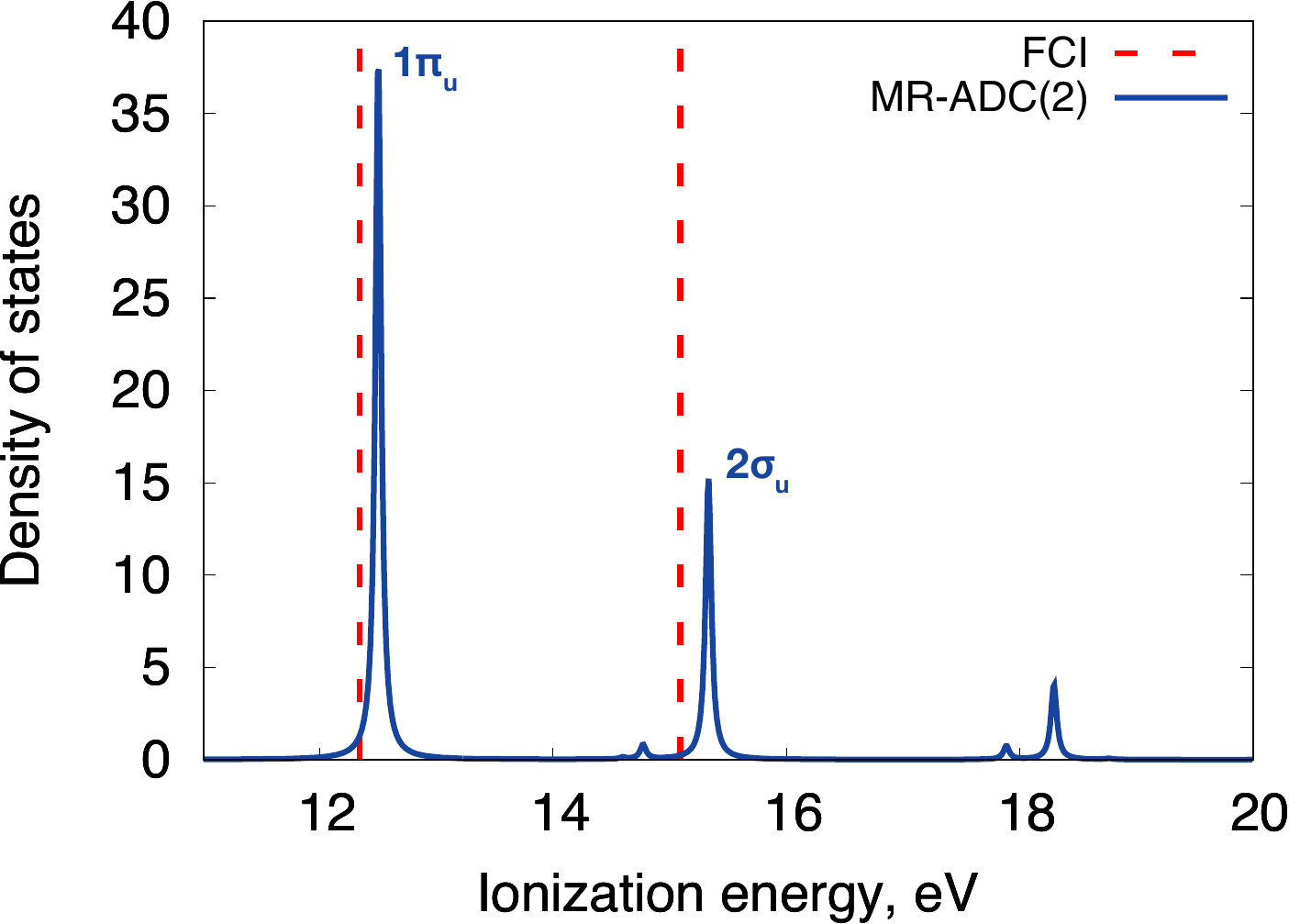}
    \captionsetup{justification=raggedright,singlelinecheck=false}
    \caption{Simulated photoelectron spectrum of carbon dimer for $r$\ce{(C-C)} = 1.2425 \angstrom computed using the MR-ADC(2) method by broadening peaks centered at vertical ionization energies with a half width of 0.03 eV (aug-cc-pVDZ basis set). Vertical dashed lines indicate FCI ionization energies for the main peaks corresponding to the $1^2\Pi_u$ and $1^2\Sigma_u^+$ states of \ce{C2+} (\cref{tab:c2}).}
    \label{fig:c2_spectrum}
\end{figure}

Next, we investigate performance of MR-ADC(2) for simulating photoelectron spectrum of \ce{C2}, which is a challenging test for {\it ab initio} methods, since electronic states of both \ce{C2} and \ce{C2+} require very accurate description of static and dynamic 
correlation.\cite{Roos:1987p399,Bauschlicher:1987p1919,Watts:199p6073,Abrams:2004p9211,Wouters:2014p1501,Holmes:2017p164111,Rosmus:1986p289,Kraemer:1987p345,Watts:1998p6073,Petrongolo:1998p4594,Ballance:2001p1201,Shi:2013p2020}
\cref{tab:c2} compares results of SR-ADC(3), MR-ADC(2), and QD-NEVPT2 with those from FCI. The MR-ADC(2) photoelectron spectrum, shown in \cref{fig:c2_spectrum}, exhibits two very intense peaks for ionizations in the $1\pi_u$ and $2\sigma_u$ orbitals, corresponding to the $1^2\Pi_u$ and $1^2\Sigma_u^+$ electronic states of \ce{C2+}, respectively. For both peaks, MR-ADC(2) is in a good agreement with FCI, showing errors in vertical ionization energies (0.16 and 0.24 eV) within \mae and \std of small molecules computed in \cref{sec:results:small_molecules}. The MR-ADC(2) results show significant improvement over SR-ADC(3), which underestimates the $1\pi_u$ and $2\sigma_u$ ionization energies from FCI by 0.65 and 1.14 eV, respectively, indicating that description of multi-reference effects is important for these ionization processes. The best agreement with FCI is demonstrated by QD-NEVPT2, with errors smaller than 0.1 eV.

\begin{figure*}[t!]
    \subfloat[]{\label{fig:h10_spectrum_short}\includegraphics[width=0.45\textwidth]{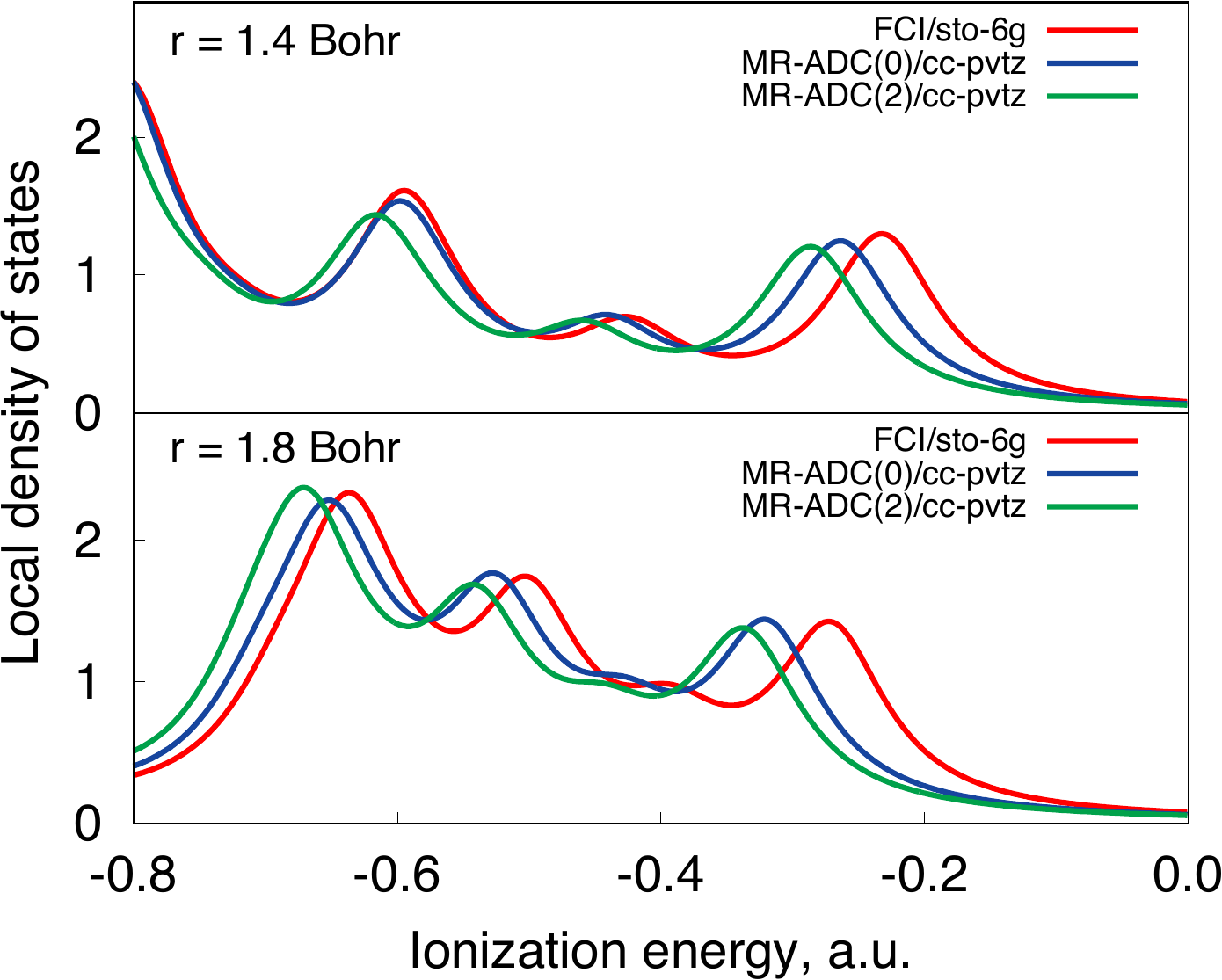}} \quad
    \subfloat[]{\label{fig:h10_spectrum_long}\includegraphics[width=0.45\textwidth]{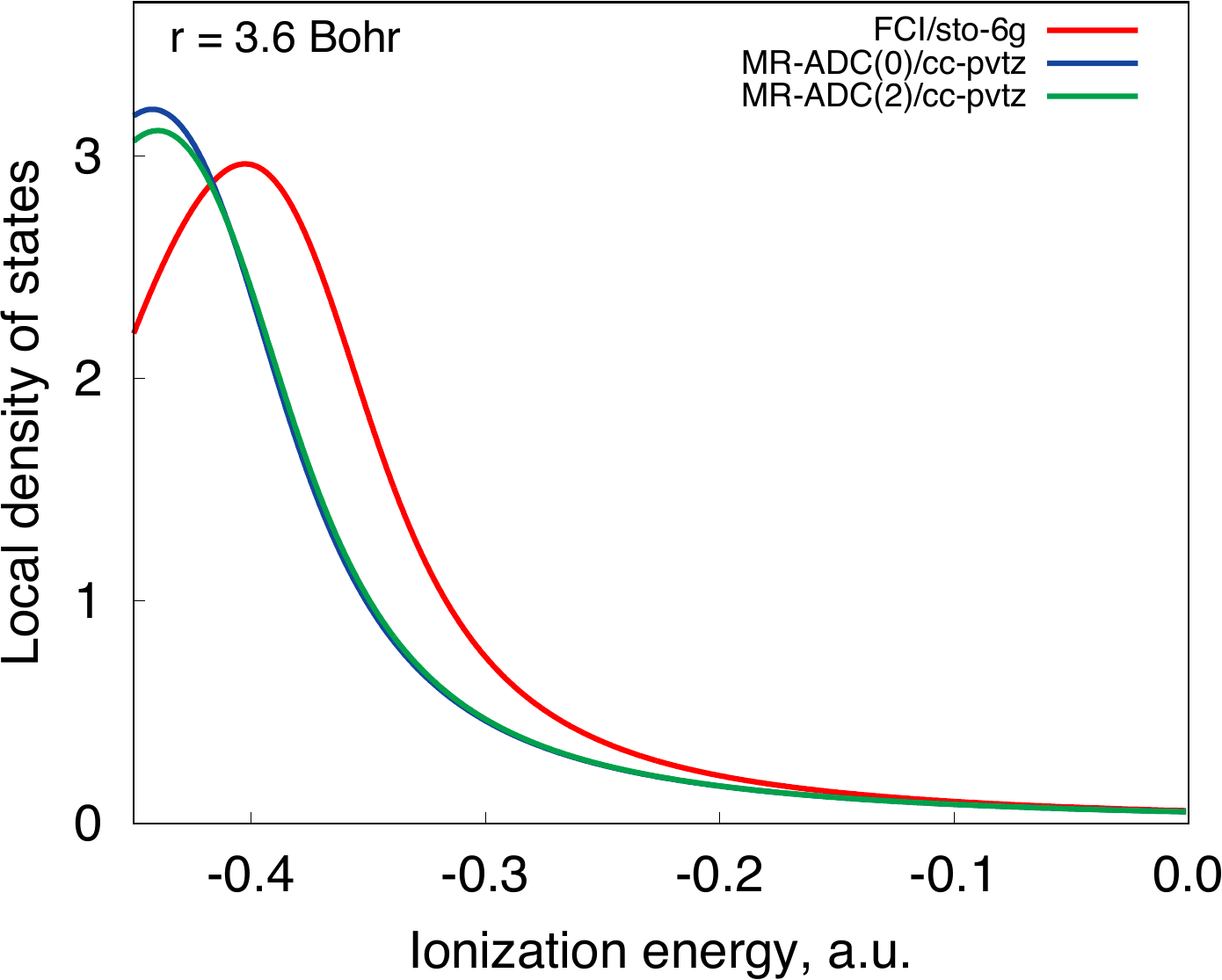}}
    \captionsetup{justification=raggedright,singlelinecheck=false}
    \caption{Local density of states (LDOS) for the equally-spaced \ce{H10} chain with $r$\ce{(H-H)} = 1.4, 1.8, and 3.6 \bohr computed using three methods with 0.05 \eh broadening. LDOS was computed at the central hydrogen atom. The MR-ADC reference CASSCF wavefunction used the (10e, 10o) active space.}
    \label{fig:h10_spectrum}
\end{figure*}

In addition to the intense peaks, the \ce{C2} photoelectron spectrum also exhibits several much weaker (satellite) peaks, which involve ionization in the $1\pi_u$ orbital accompanied by single and double $(1\pi_u)^3$ $\rightarrow$ $(3\sigma_g)^0$ excitations (\cref{tab:c2}). Out of four satellite transitions, only two are predicted by SR-ADC(3), with large errors ($>$ 2 eV). For the singly-excited shake-up states of \ce{C2+} ($1^2\Delta_g$, $1^2\Sigma_g^-$, and $1^2\Sigma_g^+$), the largest MR-ADC(2) error is 0.37 eV. However, for the doubly-excited $2^2\Pi_u$ state, MR-ADC(2) produces a larger 0.66 eV error. The QD-NEVPT2 ionization energies for all four electronic states are within 0.1 eV from the reference FCI values. The large error of MR-ADC(2) for $2^2\Pi_u$ may be attributed to the importance of differential dynamic correlation effects between this state and the ground state of \ce{C2}, since in MR-ADC(2) the first-order amplitudes of the effective Hamiltonian are preferentially determined for the latter state (\cref{sec:theory:mr_adc_ip:amplitudes}), while in QD-NEVPT2 the first-order wavefunction is constructed for each electronic state separately. The description of these differential correlation effects is expected to improve for higher-order MR-ADC approximations and will be a subject of our future research.

\subsection{Hydrogen Chains}
\label{sec:results:hydrogen_chains}

Finally, we use MR-ADC to study equally-spaced hydrogen chains \ce{H10} and \ce{H30}. Hydrogen chains are one-dimensional models for understanding strong electron correlation in molecules and materials, as well as the hydrogen phase diagram at high 
pressures.\cite{Sinitskiy:2010p014104,Stella:2011p245117,Lin:2011p096402,Tsuchimochi:2009p121102,Rusakov:2016p054106,Motta:2017p031059,Rusakov:2019p229}
An important property of a hydrogen chain is its band gap, which can be calculated as the difference between ionization potential and electron affinity. For equally-spaced chains in the thermodynamic limit, this band gap is believed to be zero at short \ce{H-H} distances ($r$), corresponding to a metallic phase, and non-zero for long distances, corresponding to an insulator. Recently, Ronca {\it et al.\@} computed local density of states (LDOS) of the H$_n$ chains ($n$ = 10, 30, and 50) at the central hydrogen atom using density matrix renormalization group (DMRG) method with the minimal STO-6G basis set.\cite{Ronca:2017p5560} They demonstrated that for near-equilibrium and stretched geometries ($r$ = 1.8 and 3.6 \bohr) LDOS converges to thermodynamic limit already for \ce{H50}, while for compressed chains ($r$ = 1.4 \bohr) finite size effects are still significant. Although in this study all valence electrons of hydrogen atoms were correlated, importance of dynamic correlation effects beyond those in the minimal one-electron basis was not investigated.

\begin{figure*}[t!]
    \subfloat[]{\label{fig:h30_spectrum_full}\includegraphics[width=0.45\textwidth]{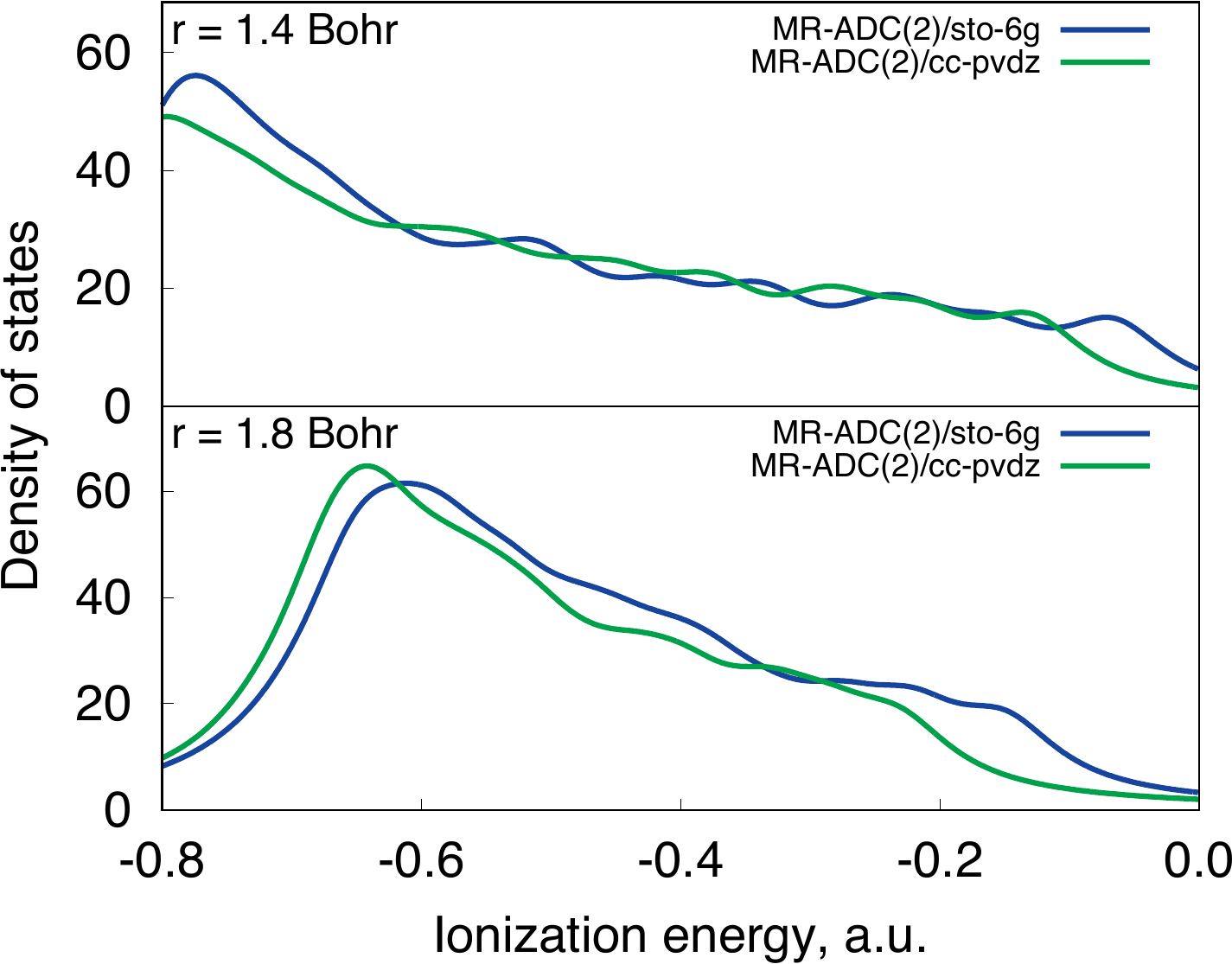}} \quad
    \subfloat[]{\label{fig:h30_spectrum_contr}\includegraphics[width=0.45\textwidth]{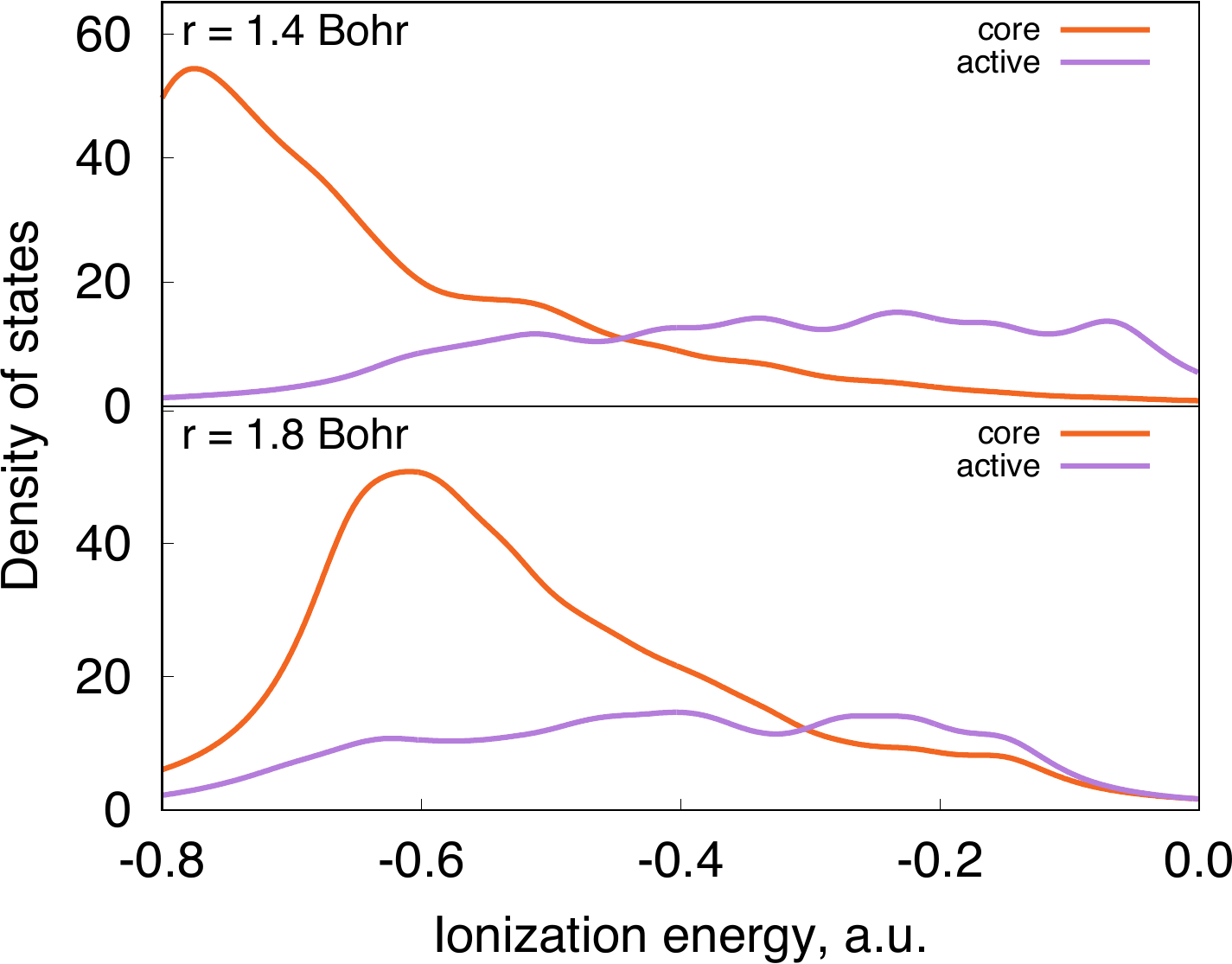}}
    \captionsetup{justification=raggedright,singlelinecheck=false}
    \caption{(a) Density of states (DOS) for the equally-spaced \ce{H30} chain with $r$\ce{(H-H)} = 1.4 and 1.8 \bohr computed using MR-ADC(2) with two basis sets and 0.05 \eh broadening. The MR-ADC(2) reference CASSCF wavefunction was computed using the (10e, 10o) active space. (b) Contributions to DOS from core and active orbitals computed using MR-ADC(2) with the STO-6G basis set.}
    \label{fig:h30_spectrum}
\end{figure*}

Here, we use MR-ADC to study effect of dynamic correlation and basis set on the density of occupied states in \ce{H10} and \ce{H30}. \cref{fig:h10_spectrum} shows LDOS of \ce{H10} for $r$ = 1.4, 1.8, and 3.6 \bohr computed at the central hydrogen atom using the MR-ADC(0) and MR-ADC(2) methods. We use the full valence (10e, 10o) active space for the CASSCF reference wavefunction and combine MR-ADC with the STO-6G\cite{Hehre:1969p2657} and cc-pVTZ basis sets, plotting LDOS for two broadening parameters: 0.05 \eh (\cref{fig:h10_spectrum}) and 0.003 \eh (Supporting Information). 
For the minimal STO-6G basis, results of MR-ADC(0) and MR-ADC(2) are equivalent to FCI. The computed LDOS are in a very good agreement with LDOS obtained by Ronca {\it et al.\@} employing the dynamical DMRG algorithm for all three geometries.\cite{Ronca:2017p5560}
Next, we consider LDOS computed using MR-ADC(0) with the larger cc-pVTZ basis set. Increasing the basis set shifts LDOS to higher ionization energies, relative to LDOS from FCI/STO-6G. For short bond distances ($r$ = 1.4 and 1.8 \bohr), the largest shifts are observed for the lowest-energy peaks corresponding to the ionization potential of the system ($\sim$ 0.03 and 0.05 \eh, respectively). For the stretched chain ($r$ = 3.6 \bohr), increasing the basis set compresses LDOS and shifts the position of its maximum by $\sim$ 0.04 \eh. Incorporating dynamic correlation effects from MR-ADC(0) to MR-ADC(2) shifts the computed LDOS further to lower energies. For most of the peaks at short bond distances, the computed shifts are $\le$ 0.02 \eh. For $r$ = 3.6 \bohr, including dynamic correlation does not significantly change position of the first band in the spectrum. Overall, our results suggest that increasing the one-electron basis set and incorporating dynamic correlation effects are similarly important and should be both taken into account in accurate computations of LDOS for hydrogen chains.

An attractive feature of MR-ADC is that it is not limited to describing ionization processes only in active orbitals. We demonstrate this by computing total density of occupied states (DOS) for the \ce{H30} chain using MR-ADC(2) with the \mbox{(10e, 10o)} active space. Since for this system we do not include all valence orbitals in the active space, we do not consider the stretched $r$ = 3.6 \bohr geometry. \cref{fig:h30_spectrum_full} shows MR-ADC(2) DOS computed using the STO-6G and cc-pVDZ basis sets. For both geometries, DOS computed using MR-ADC(2) with the STO-6G basis closely resembles LDOS of the same system from the DMRG study of Ronca {\it et al}.\cite{Ronca:2017p5560} \cref{fig:h30_spectrum_contr} plots contributions to MR-ADC(2)/STO-6G DOS from core and active orbitals separately. For the compressed chain ($r$ = 1.4 \bohr), contributions from active orbitals dominate the low-energy part of the spectrum, whereas, for equilibrium geometry ($r$ = 1.8 \bohr), core and active orbitals have similar contributions to DOS already for low ionization energies. Increasing the basis set from STO-6G to cc-pVDZ shifts peaks in DOS to higher energies. As for the \ce{H10} chain, the largest shifts are observed for the peak at the first ionization potential.

\section{Conclusions}
\label{sec:conclusions}

We presented derivation and implementation of second-order multi-reference algebraic diagrammatic construction theory (MR-ADC(2)) for simulating ionization energies and transition properties of strongly correlated systems. In MR-ADC(2), ionization energies and spectral properties are determined from poles and residues of the one-electron Green's function that is evaluated to second order in multi-reference perturbation theory with respect to a complete active space (CAS) reference wavefunction. In contrast to conventional second-order multi-reference perturbation theories (such as multi-state CASPT2 or NEVPT2), MR-ADC(2) describes ionization in all orbitals (e.g., core and active), does not require using state-averaged wavefunctions to compute higher-energy ionized states, and provides direct access to spectroscopic properties. Although equations of MR-ADC(2) depend on four-particle reduced density matrices, we demonstrated that computation of these large matrices can be completely avoided by constructing efficient intermediates, without introducing any approximations. The resulting MR-ADC(2) implementation has a lower $\mathcal{O}(N_{\mathrm{det}} N^6_{\mathrm{act}})$ computational scaling with respect to the number of active orbitals ($N_{\mathrm{act}}$), compared to the $\mathcal{O}(N_{\mathrm{det}} N^8_{\mathrm{act}})$ scaling of conventional multi-reference perturbation theories. 

We benchmarked accuracy of MR-ADC(2) for predicting ionization energies of eight small molecules, carbon dimer (\ce{C2}), and hydrogen chains (\ce{H10} and \ce{H30}), against results from full configuration interaction (FCI). For small molecules, MR-ADC(2) shows consistent performance for equilibrium and stretched geometries, with mean absolute errors of $\sim$ 0.5 eV in ionization energies and 0.1 eV in energy separations between ionized states. For \ce{C2}, MR-ADC(2) predicts energies of the main and singly-excited satellite peaks within 0.4 eV from the FCI reference values, but has a large $\sim$ 0.7 eV error for the doubly-excited satellite transition. The QD-NEVPT2 method shows smaller ($\sim$ 0.1 eV) errors than MR-ADC(2) for all ionized states of \ce{C2}, providing an improved description of differential dynamic correlation effects, which are important for this system. We expect that these effects will be better described using the higher-order MR-ADC approximations, which will be one of the directions of our future work. Finally, we used MR-ADC(2) to investigate density of occupied states (DOS) in \ce{H10} and \ce{H30}. For \ce{H10}, our results provide numerical evidence that including dynamic correlation effects beyond those incorporated in a full valence CAS and increasing single-particle basis set have a similar effect on the computed local DOS. Since dynamic correlation is a local phenomenon, we expect that its effect will be similar for longer hydrogen chains as well. For \ce{H30}, we showed that DOS computed using MR-ADC(2) combined with a small (10e, 10o) active space is in a very good agreement with previously reported results from density matrix renormalization group, incorporating 30 electrons and orbitals in the active space.

Overall, our results suggest that MR-ADC is a promising theoretical approach for computing ionization energies and spectral densities of multi-reference systems and encourage its further development. Future work will be directed towards more efficient implementation of MR-ADC(2) for systems with a large number of electrons and active orbitals, as well as the development of more accurate MR-ADC approximations that will incorporate description of higher-order dynamic correlation effects. We also plan extending our MR-ADC methods to simulations of core-level ionizations in X-ray photoelectron spectroscopy, which has become a widely used tool for experimental investigations of molecules and materials.

\acknowledgement
This work was supported by start-up funds provided by the Ohio State University. Computations were performed at the Ohio Supercomputer Center under projects PAS1317 and PAS1583.\cite{OhioSupercomputerCenter1987} The authors would like to thank Sandeep Sharma for useful suggestions about carrying out SHCI computations.

\suppinfo
Comparison of MR-ADC(2) results with exact and approximate second-order amplitudes. Equations of the MR-ADC(2) method for the $\mathbf{M}$, $\mathbf{T}$, and $\mathbf{S}$ matrices. 
Density of states for \ce{H10} with a small broadening parameter.

\providecommand{\latin}[1]{#1}
\makeatletter
\providecommand{\doi}
  {\begingroup\let\do\@makeother\dospecials
  \catcode`\{=1 \catcode`\}=2 \doi@aux}
\providecommand{\doi@aux}[1]{\endgroup\texttt{#1}}
\makeatother
\providecommand*\mcitethebibliography{\thebibliography}
\csname @ifundefined\endcsname{endmcitethebibliography}
  {\let\endmcitethebibliography\endthebibliography}{}

\end{document}